\documentclass[preprint,11pt,authoryear]{elsarticle}
\usepackage{amssymb}
\usepackage{amsthm}
\usepackage{setspace}
\usepackage{epsfig}
\usepackage{color}
\usepackage{amsthm,amsmath,amssymb}
\usepackage{mathrsfs}
\usepackage{bbm}
\usepackage{amssymb}
\usepackage{amsmath}
\usepackage{amsfonts}
\usepackage{natbib}
\usepackage{verbatim}
\usepackage{graphicx}
\usepackage{geometry}
\newtheorem{proposition}{Proposition}[section]
\newtheorem{corollary}{Corollary}[section]
\newtheorem{remark}{Remark}[section]
\newtheorem{definition}{Definition}[section]
\newtheorem{example}{Example}[section]
\newtheorem{lemma}{Lemma}[section]

\journal{ Insurance: Mathematics and Economics}
\begin{document}
	
	\begin{frontmatter}
		\title{Dynamic star-shaped risk measures and  $g$-expectations}
		\author[math]{Dejian Tian\corref{correspondingauthor}}
		\ead{djtian@cumt.edu.cn}
		\author[math]{Xunlian Wang}
		\ead{xlwang@cumt.edu.cn}
		\address[math]{School of Mathematics, China University of Mining and Technology, Xuzhou, P.R. China}
		\cortext[correspondingauthor]{Corresponding author}
				
		\begin{abstract}
Motivated by static monetary and star-shaped risk measures, this paper investigates representation theorems within a dynamic framework. We show that dynamic monetary risk measures can be represented as the lower envelope of a family of dynamic convex risk measures,  and normalized dynamic star-shaped risk measures can be represented  as that of a family of normalized dynamic convex risk measures.  Furthermore, we establish the intrinsic link between dynamic monetary risk measures and dynamic star-shaped risk measures. Moreover, we examine the associated sensitivity and time consistency issues for such risk measures.  Finally, we characterize star-shapedness of risk measures induced by bounded quadratic BSDE solutions and illustrate normalized time-consistent examples through $g$-expectations.
\end{abstract}
		
		\begin{keyword}
dynamic monetary risk measure\sep dynamic star-shaped risk measure\sep $g$-expectation\sep  representation theorem
		\end{keyword}
		
	\end{frontmatter}
	
	%% \linenumbers
	
	%% main 
	
	\section{Introduction}

The theory of risk measures is a fruitful research area in the field of mathematical finance.    Axiomatically based risk measures have been extensively  studied  because most axioms possess desirable economic characteristics.  In their seminal work,  from an axiomatic  perspective of capital requirements,  \cite{ADEH1999} proposed the concept of coherent risk measures which satisfy  monotonicity,  translation-invariance, subadditivity,  and positive homogeneity.   Further,  by substituting  convexity for subadditivity and positive homogeneity,   \cite{FS2002}  and \cite{FRG2002} introduced a more general concept of convex risk measures.    Since then, convex and coherent risk measures have garnered significant attention.

Despite the theoretical appeal of convex risk measures, the convexity axiom may be too restrictive in the presence of market frictions such as illiquidity, transaction costs, or price impact. In such environments, the risk of a scaled position can be more than proportional to its size, a phenomenon naturally captured by the \textit{star-shaped} property $\rho(\lambda X) \ge \lambda \rho(X)$ for all $\lambda \ge 1$. This condition arises in models of granular price systems (\cite{CDFM2015}) and reflects the increasing marginal risk typical of liquidity shortages. Moreover, the envelope approach is related to the representation
of positively homogeneous functions by sublinear majorants
(\cite{C2000}), while the connection between star-shaped risk
measures and quasi-logconvex preferences is studied in
\cite{LR2022}. By encompassing both normalized convex risk measures and popular non-convex examples such as Value at Risk, the class of star-shaped risk measures offers a flexible yet economically meaningful framework for risk assessment.

Motivated by the need for non-convex risk measures, two parallel strands of research have recently emerged in the static setting. On one hand, \cite{JXZ2020} studied \textit{monetary risk measures}, those satisfying only monotonicity and translation invariance, and proved that any static monetary risk measure can be represented as the lower envelope of a family of convex risk measures. On the other hand, \cite{CCMW2022} introduced \textit{star-shaped risk measures} and established  that a normalized static star-shaped risk measure admits a similar representation via normalized convex risk measures. Subsequently, \cite{MR2022} uncovered the precise relationship between these two classes, showing that under mild conditions, a monetary risk measure is only a translation away from star-shapedness.  All these results, however, pertain to the static setting. A natural question follows: do these representation theorems remain valid in a dynamic framework where information evolves over time? This is the central question addressed in the present paper.

To answer this question, we first recall the framework of dynamic risk measures. Dynamic coherent and convex risk measures were introduced by \cite{Riedel2004}, \cite{FRG2004} and \cite{DS2005}; see also \cite{FP2006}, \cite{APDFEJHK2007}, \cite{KS2007}, \cite{BN2008,BN2009} and \cite{FS16}. A powerful tool for constructing dynamic risk measures is the theory of $g$-expectations, pioneered by \cite{P1997} via backward stochastic differential equations (BSDEs). The connection between $g$-expectations and risk measures has been extensively studied; see \cite{Rosazza Gianin2006}, \cite{Jiang2005,Jiang2008}, \cite{HMPY2008}, \cite{DPRG2010} and \cite{Ji2019}. In this paper, we will employ this BSDE framework to construct explicit examples of dynamic star-shaped risk measures.

Our main contributions are briefly summarized as follows. First, we show that dynamic monetary risk measures can be represented as attained lower envelopes of dynamic convex risk measures at each fixed time, and normalized dynamic star-shaped risk measures admit such representations with normalized convex components. Specifically, a dynamic monetary risk measure $\rho_{\cdot}$ admits
$$
\rho_t(X)=\mathop{\operatorname{ess}\inf}_{\lambda\in\Lambda}
\mathop{\operatorname{ess}\sup}_{Q\in\mathcal D_t}
\bigl(E_Q[-X\mid\mathscr F_t]-\alpha_{t,\lambda}(Q)\bigr),
\qquad t\in[0,T],
$$
where $\mathcal D_t$ consists of probabilities $Q\ll P$ satisfying $Q|_{\mathscr F_t}=P|_{\mathscr F_t}$, and the penalties are conditionally convex under normalized density mixtures; see Proposition \ref{prop:3-1}. Proposition \ref{prop:3-2} provides the normalized star-shaped counterpart. Corollary  \ref{cor:3-1} establishes an attained lower envelope representation for dynamic positively homogeneous risk measures by dynamic coherent majorants, without requiring continuity from above or countably additive dual representations for those components.

Although the main construction technique for convex sets employed in proving the representation theorem has been utilized in the literature (\cite{JXZ2020}; \cite{CCMW2022}), the elaborate properties of the constructed sets require  a more nuanced analysis in the dynamic framework. For example, we use the Koml\'{o}s-type convergence result  to establish the weak$*$-closed property of $\mathscr{A}_{t}(Z)$  in the context of dynamic star-shaped risk measures.
	
Second, we establish the relationship between dynamic monetary risk measures and dynamic star-shaped risk measures.  Proposition \ref{prop:3-5} shows that  dynamic monetary risk measures are  merely a translation of dynamic star-shaped risk measures at a specific position, under some suitable conditions.  
The dynamic setting requires additional care with conditional
structure and essential suprema and infima, extending the
static analysis of \cite{MR2022}.  Additionally, we further investigate the sensitivity and time consistency properties of these risk measures.

Third, we consider a specific class of time consistent dynamic star-shaped risk measures induced by $g$-expectations and backward stochastic differential equations (BSDEs).  By leveraging the theory and tools of BSDEs, under conditions (C2)-(C3) in the bounded solution quadratic BSDE setting,  Proposition \ref{prop:4-1}  establishes the triple equivalence  that  $\rho^{g}(\cdot)$ is a  normalized  static star-shaped risk measure if and only if  $\{\rho_t^{g}(\cdot)\}_{0\leq t\leq T}$ is a 
normalized time consistent  dynamic  star-shaped risk measure if and only if the generator $g$ is star-shaped in  its $z$-argument. 
Several concrete examples, including $\alpha$-maxmin expectations and entropic risk measures with drift ambiguity, are also presented. The latter admit an optimistic selection interpretation and provide nonconvex time consistent star-shaped risk measures.

We end this introduction with an overview of closely related literature on star-shaped risk measures. \cite{Righi2022} introduced static star-shaped acceptability indexes, \cite{NTJ2025} analyzed set-valued star-shaped risk measures and  \cite{LRZ2024}  developed law-invariant return and star-shaped risk measures.  In concurrent and independent work, \cite{LRM2025} established related BSDE representation results in a Lipschitz setting and also study minimal supersolutions.  Our work complements
these results through fixed time conditional envelope
representations, the monetary and star-shaped link, and a
characterization within the bounded quadratic BSDE setting.

The paper is structured as follows.  Section \ref{sec:2} presents axiomatic conditions for dynamic risk measures and properties of their acceptance sets; Section \ref{sec:3} establishes representation theorems for dynamic monetary and normalized dynamic star-shaped risk measures, derives their transformational relationship, and investigates their sensitivity and time consistency; Section \ref{sec:4} constructs and analyzes time consistent dynamic star-shaped risk measures induced by $g$-expectations; finally, Section \ref{sec:5}  concludes the paper.

\section{Preliminaries}\label{sec:2}

We consider a probability space $(\Omega, \mathscr{F}, P)$ with a filtration $\mathbb F= \left(\mathscr{F}_{t}\right)_{0 \leq t \leq T} $ satisfying the usual conditions.  $ T \in(0, \infty) $  denotes a fixed time horizon, where we impose  that $ \mathscr{F}=\mathscr{F}_{T} $ and $ \mathscr{F}_{0}=\{\emptyset, \Omega\}$.  For $ t\in [0,T]$,  $L_{t}^{\infty}(P):=L^{\infty}\left(\Omega, \mathscr{F}_{t}, P\right) $ denotes the space of all essentially bounded $ \mathscr{F}_{t} $-measurable random variables and $L_{t,+}^{\infty}(P)$ denotes its positive cone,  
and $ L_{t}^{0}((-\infty,+\infty])$ is the set of all $\mathscr{F}_{t}$-measurable mappings $\Omega\rightarrow (-\infty,+\infty]$. The positive cone of  $L_{t}^{1}(P)=L^{1}\left(\Omega, \mathscr{F}_{t}, P\right) $ is denoted by $L_{t,+}^{1}(P)$.   In particular, $\|X\|_\infty=\operatorname{ess\,sup}_P|X| $.
All equalities and inequalities throughout the paper should be understood as holding  $P$-almost surely, unless otherwise specified.
	
Denote $\mathscr{M}_{1}(P):=\mathscr{M}_{1}(\Omega, \mathscr{F}, P)$ as the set of all probability measures on  $(\Omega, \mathscr{F})$  that are absolutely continuous with respect to  $P$.  For each $t\in[0,T]$,  let 
$\mathscr {Q}_{t}:=\left\{Q\in\mathscr{M}_{1}(P) \mid Q \sim P ~\text{on}~ \mathscr{F}_t \right\}$, and  denote $\mathcal{M}_{1}^{e}(P):=\mathscr {Q}_{T}$.  
For conditional dual representations,  we will also use the following classes
\begin{align*}
\mathcal D_t=\{Q\in\mathscr M_1(P):Q|_{\mathscr F_t}=P|_{\mathscr F_t}\}  \text{~~and~~}  \mathcal L_t=\{L\in L_{T,+}^1(P):E_P[L\mid\mathscr F_t]=1\}.
\end{align*}
Obviously,  $\mathcal D_t\subset\mathscr Q_t$, $\mathcal D_T=\{P\}$ and elements $Q\in\mathcal D_t$ are in bijective correspondence with $L = \mathrm{d}Q/\mathrm{d}P\in\mathcal L_t$.   For $Q\in\mathcal D_t$, with density $L=dQ/dP\in\mathcal L_t$, and bounded $H$,
$E_Q[H\mid\mathscr F_t]=E_P[LH\mid\mathscr F_t].$

Building on the static (unconditional) framework of risk measures, we study dynamic risk measures, which are mappings from $L_{T}^{\infty}(P)$ to  the space of $\mathbb F$-adapted stochastic processes,  where $L_{T}^{\infty}(P)$  is understood as the set of discounted terminal values of financial positions. This section  collects some preliminary definitions and key properties of dynamic risk measures.  For further details, readers may consult Chapter 4 of \cite{FS16} and \cite{CCMW2022} for the static case, and Chapter 11 of \cite{FS16} for the dynamic setting.

 A mapping $\rho_{t}:L_{T}^{\infty}(P)\rightarrow L_{t}^{\infty}(P) $ for each $t\in[0,T]$  may satisfy the following axioms:
	\begin{itemize}
		\item [(A1)] Monotonicity: $\rho_{t}(X)\geq \rho_{t}(Y)$ for all  $X, Y\in L_{T}^{\infty}(P)$ with $X\leq Y$.
		\item[(A2)] Conditional translation invariance: $\rho_{t}(X+Y_{t})=\rho_{t}(X)-Y_{t}$, for all $X\in L_{T}^{\infty}(P)$ and $Y_{t}\in L_{t}^{\infty}(P)$.
		\item [(A3)] Normalization: $\rho_{t}(0)=0$.
		\item [(A4)] Conditional convexity: $\rho_{t}(\alpha X+(1-\alpha)Y)\leq \alpha\rho_{t}(X)+ (1-\alpha)\rho_{t}(Y)$ for every pair $X, Y\in L_{T}^{\infty}(P)$ and all $\alpha \in L_{t}^{\infty}(P)$ with $0\leq \alpha \leq 1.$
		\item [(A5)] Conditional positive homogeneity: $\rho_{t}(\alpha X)= \alpha\rho_{t}(X)$ for  all  $X\in L_{T}^{\infty}(P)$ and $\alpha \in L_{t}^{\infty}(P)$ with $\alpha \geq0$.
		\item [(A6)] Conditional star-shapedness: $\rho_{t}(\alpha X)\geq \alpha\rho_{t}(X)$ for  all  $X\in L_{T}^{\infty}(P)$ and $\alpha \in L_{t}^{\infty}(P)$ with $\alpha \geq1$.		
	\end{itemize}

\begin{definition}\label{df:2.1.}
	A  family
	$\rho=\{\rho_{t}\}_{t\in[0,T]}$  is called a dynamic monetary risk measure if $ \rho$ satisfies (A1) and (A2);  $\rho$ is called a normalized dynamic star-shaped risk measure if $\rho$ fulfills (A1)-(A3) and (A6);  
$\rho$ is called a dynamic positively homogeneous risk measure if $\rho$ satisfies (A1)-(A2) and (A5).   Furthermore, $\rho$ is called a dynamic convex risk measure if $\rho$ satisfies (A1)-(A2) and (A4),  and $\rho$ is called a dynamic coherent risk measure if $\rho$ satisfies (A1)-(A2) and (A4)-(A5). 
\end{definition}

For monetary risk measures, star-shapedness without the qualifier “normalized” means condition (A6), without imposing (A3).   This implies $\rho_t(0)\leq0$, but equality need not hold. We also refer to each individual mapping $\rho_{t}$ as a dynamic risk measure (at time 
$t$),  whenever the family $\rho=(\rho_{t})_{t\in[0,T]}$ satisfies the corresponding axioms.  When $t=0$, the mapping $\rho_{0}$ reduces to the static version of each respective definition.

Given a dynamic monetary risk measure  $ \rho$, its  acceptance set is defined as
$$ \mathscr{A}_{t}:=\left\{X \in L_{T}^{\infty}(P) \mid \rho_{t}(X) \leq 0\right\},  \quad t\in[0,T].$$
Additionally, given a non-empty set $ \mathscr{A}_{t}\subset  L_{T}^{\infty}(P)$, let the mapping $ \rho_{\mathscr{A}_{t}} $  be defined by
	\begin{equation}\label{accp} \rho_{\mathscr{A}_{t}}(X):=\operatorname{ess} \inf \left\{m\in L_{t}^{\infty}(P) \mid X+m \in \mathscr{A}_{t}\right\}, \quad X\in L_{T}^{\infty}(P) .\end{equation}
	 A non-empty set $ \mathscr{A}_{t}\subseteq  L_{T}^{\infty}(P)$ may possess  the following properties:
	\begin{itemize}
		\item [(B1)] Solidity: $ \mathscr{A}_{t} $ is solid if  $X\in\mathscr{A}_{t} $ and  $Y\in L_{T}^{\infty}(P)$  with $Y\geq X $  implies $ Y\in  \mathscr{A}_{t}$.
		\item[(B2)] Monetarity: $ \mathscr{A}_{t} $ is monetary  if  $\operatorname{ess} \inf (L_{t}^{\infty}(P)\cap \mathscr{A}_{t})\in L_{t}^{\infty}(P).$
		\item [(B3)] Normalization: $ \mathscr{A}_{t} $ is normalized  if  $\operatorname{ess} \inf (L_{t}^{\infty}(P)\cap \mathscr{A}_{t})=0. $
		\item [(B4)] Conditional conicity: $ X \in \mathscr{A}_{t} $ implies $ \alpha X \in \mathscr{A}_{t} $ for  all $\alpha \in L_{t}^{\infty}(P)$ with $\alpha \geq 0.$
		\item [(B5)] Conditional convexity: $ X, Y\in \mathscr{A}_{t} $ implies $ \alpha X+(1-\alpha)Y \in \mathscr{A}_{t} $ for  all $\alpha \in L_{t}^{\infty}(P)$ with $0\leq \alpha \leq 1.$
		\item [(B6)] Conditional star-shapedness: $\mathscr{A}_{t}$ is conditionally  star-shaped at $ \mathscr{B}\subseteq L_{T}^{\infty}(P) $ if $X\in \mathscr{A}_{t}$  implies that $\alpha X+(1-\alpha)Y \in \mathscr{A}_{t} $ for all $ Y\in \mathscr{B} $ and  all $\alpha \in L_{t}^{\infty}(P)$ with $0\leq \alpha \leq 1$.  If $ \mathscr{A}_{t} $ is conditionally star-shaped at a singleton  $ \mathscr{B}=\left\{Y\right\} $, we say that $ \mathscr{A}_{t} $ is  conditionally star-shaped at $ Y. $ Unless otherwise specified, conditional star-shapedness is  understood as conditionally star-shaped at $0$.	
		\item[(B7)] Conditional decomposability: $1_AX+1_{A^c}Y\in\mathscr A_t$ for  $X,Y\in \mathscr{A}_{t} $ and $A\in\mathscr F_t$.
\end{itemize}
 
Normalization anchors the capital requirement of the zero position at zero.  For any accepted position, the associated capital requirement is nonpositive.  The acceptance set $\mathscr{A}_t$ consists precisely of those positions that need no additional capital at  time $t$.  In the dynamic framework, normalization must hold for every $t\in[0,T]$,  thereby anchoring the risk assessment at zero across all information stages.

Define
$$\|X\|_t=\operatorname{ess} \inf \{m\in L_t^\infty(P):  |X|\leq m\}.$$
Conditional-norm closedness means that $X_n\in\mathscr A_t$, $X\in L_T^\infty(P)$, and $\|X_n-X\|_t\to0$ almost surely imply $X\in\mathscr A_t$.  The following lemma summarizes the fundamental relationship between dynamic risk measures and acceptance sets.

\begin{lemma}(\cite{KS2007}, Lemmas 3.1 and 3.2)\label{lem:5-1}
Given a dynamic monetary risk measure  $\rho_{t}$ for each $t\in[0,T]$,   its acceptance set $\mathscr{A}_{\rho_{t}}$  
is non-empty, and satisfies (B1)-(B2),  (B7), and is conditionally-norm closed, with $\operatorname{ess} \inf (L_{t}^{\infty}(P)\cap \mathscr{A}_{\rho_{t}})=\rho_t(0)$.   Conversely,  for  any $t\in[0,T]$,  and a nonempty subset $ \mathscr{A}_{t} \subset L_T^{\infty}(P) $ satisfying  (B1), (B2) and (B7), 
the mapping $ \rho_{\mathscr{A}_{t}}$ defined by   \eqref{accp} is a dynamic monetary risk measure.   If $ \mathscr{A}_{t} $ is also conditionally-norm closed, then  $\mathscr A_t=\mathscr A_{\rho_{\mathscr A_t}}$.

For each fixed $t$, let $\rho_t$ be a dynamic monetary risk measure and $\mathscr{A}_{\rho_t}$ its acceptance set.  Then the following equivalences hold:
\begin{itemize}
\item[(i)]  $\rho_t$  is a dynamic  star-shaped risk measure  iff  $ \mathscr{A}_{\rho_t} $  satisfies   (B6) at zero.		
\item[(ii)]  $\rho_t$ is a dynamic convex risk measure iff   $\mathscr{A}_{\rho_t} $  satisfies  (B5).		
\item[(iii)] $\rho_t$  is a dynamic coherent risk measure iff  $ \mathscr{A}_{\rho_t} $ satisfies  (B5) and  (B4). 	
\item[(iv)] $\rho_t$ is dynamic positively homogeneous risk measure if and only if $ \mathscr{A}_{\rho_t} $ satisfies  (B4).
\item[(v)] $\rho_t$ is normalized risk measure if and only if $ \mathscr{A}_{\rho_t} $ satisfies (B3).	
\end{itemize}		
\end{lemma}
\begin{proof}
The dynamic convex and coherent correspondences can be obtained from Lemmas 3.1 and 3.2 in \cite{KS2007},  after applying the sign transformation $\Phi_t=-\rho_t$. Their conditional regularity condition is restated as  (B7) here. To encompass dynamic monetary risk measure without the convexity property, to facilitate readers' understanding, we provide the essential reconstruction reasoning, followed by extending to the  star-shaped cases.

For a dynamic monetary risk measure, it is known that (A1) and (A2) imply that  for any $X,Y\in  L_{T}^{\infty}(P)$, 
$|\rho_t(X)-\rho_t(Y)|\leq\|X-Y\|_t$ and  $\rho_t(1_AX+1_{A^c}Y)
=1_A\rho_t(X)+1_{A^c}\rho_t(Y)$ for $A\in\mathscr{F}_t.$ 
This gives (B7) and  conditional-norm closedness of its acceptance set. 
(B1) follows from monotonicity and $\operatorname{ess} \inf (L_{t}^{\infty}(P)\cap \mathscr{A}_{\rho_{t}})=\rho_{t}(0)\in \mathscr{A}_{\rho_{t}}.$   Moreover,  $\rho_{\mathscr A_{\rho_t}}(X)=\operatorname{ess\,inf}\{m\in L_t^\infty(P):\rho_t(X)-m\le0\}=\rho_t(X).$

Conversely, let $\mathscr{A}_t\neq\emptyset$ and  satisfy (B1), (B2), (B7), and 
$\rho_{\mathscr{A}_t}(X)=\operatorname{ess} \inf \{m\in L_t^\infty(P):X+m\in\mathscr{A}_t\}.$ Suppose $Z\in \mathscr{A}_t$, then by (B1) and (B2), one can verify that $\operatorname{ess} \inf (L_{t}^{\infty}(P)\cap \mathscr{A}_{t})-||X||_{\infty}\leq \rho_{\mathscr{A}_t}(X)\leq ||X-Z||_{\infty}$. Hence, it is a finite mapping for all $X\in L_{T}^{\infty}(P)$. (A1) and (A2) are also easily obtained, which means $\rho_{\mathscr{A}_t}$ is a dynamic monetary risk measure. 

For any $m,n\in L^{\infty}_{t}(P)$ such that $X+m$ and $X+n$ are in $\mathscr{A}_t$, by (B7), one can verify that $X+m\wedge n \in\mathscr{A}_t$, which means the  set is downward directed.  Thus, for the given $X$, there exists a sequence $m_{n}\in L^{\infty}_{t}(P)$ such that $X+m_{n} \in\mathscr{A}_t$ and $m_{n}\downarrow \rho_{\mathscr{A}_t}(X)$. If $ \mathscr{A}_{t} $ is  conditionally-norm closed, then $X+\rho_{\mathscr{A}_t}(X)\in \mathscr{A}_t$.  Consequently,  $X\in\mathscr{A}_t$ implies $\rho_{\mathscr{A}_t}(X)\leq0$, i.e., $X\in \mathscr{A}_{\rho_{\mathscr{A}_t}}$, and the converse follows from solidity. This proves  $\mathscr{A}_t=\mathscr{A}_{\rho_{\mathscr{A}_t}}$. Noting that these above facts do not require convexity.

Cases (ii) and (iii) are covered by \cite{KS2007}.  Cases (iv) and (v) are also easily obtained. We only verify case (i).  If $\mathscr{A}_{\rho_{t}}$
 is conditionally star-shaped at zero, apply (B6) to $X+\rho_{t}(X)$.  For $k\in L_t^\infty(P)$ with $0\leq k\leq1$, it implies
$kX+k\rho_{t}(X)\in\mathscr{A}_{\rho_{t}}$ and  $\rho_{t}(kX)=\rho_{\mathscr{A}_{\rho_{t}}}(kX)\leq k\rho_{t}(X)$. Therefore, $\rho_{t}(aX)\geq a\rho_{t}(X)$ for $a\in L_t^\infty(P)$ with $a\geq1$. 

Conversely, assume $\rho_t$  is a dynamic  star-shaped risk measure. For $0\leq k\leq1$ in $L_t^\infty(P)$, set $k_n=k\vee(1/n)$.  Applying (A6) to $k_nX$ with the bounded multiplier $1/k_n$ gives
$\rho_{t}(k_nX)\leq k_n\rho_{t}(X)$. 

Since $\|(k_n-k)X\|_\infty\leq\|X\|_\infty/n$,
Lipschitz continuity gives
$\rho_t(kX)\leq k\rho_t(X)$.
Thus, $X\in\mathscr A_{\rho_t}$ implies
$kX\in\mathscr A_{\rho_t}$, proving (B6) at zero. \hfill$\Box$
\end{proof}

\section{Representations of dynamic monetary risk measures and dynamic star-shaped risk measures}\label{sec:3}
	
%Inspired by the static results of \cite{JXZ2020}  and \cite{CCMW2022},  this section devotes to establishing the representation theorems for dynamic monetary risk measures and normalized dynamic star-shaped risk measures.  Similar to \cite{MR2022},   we obtain the relationship between dynamic monetary risk measures and dynamic star-shaped risk measures.   Besides, the sensitivity properties of  dynamic monetary (star-shaped) risk measures are also discussed. 

Motivated by  the static results of \cite{JXZ2020} and \cite{CCMW2022}, this section  is devoted to establishing  representation theorems for dynamic monetary risk measures and normalized dynamic star-shaped risk measures.  Following the approach of  \cite{MR2022}, we  establish the precise relationship  between dynamic monetary risk measures and dynamic star-shaped risk measures. In addition, we  investigate  the sensitivity properties of these dynamic risk measures.
	
\subsection{Representations of dynamic monetary risk measures}

The following proposition shows that a dynamic monetary risk measure  can be represented as  the lower envelope of a family of dynamic convex risk measures.	
	
\begin{proposition}\label{prop:3-1}
For a mapping $\rho_{t}:L_{T}^{\infty}(P)\rightarrow L_{t}^{\infty}(P)$,  $t\in[0,T]$, the following assertions are equivalent.	
		
(a) $ \rho_{t} $ is a dynamic monetary risk measure. 
		
(b) There exists a nonempty family $\left\{\alpha_{t,\lambda} \mid \lambda \in \Lambda\right\}$ of conditionally convex functionals $\alpha_{t,\lambda}:\mathcal{D}_{t}\rightarrow L^{0}_{t}((-\infty, \infty])$ with 
$\mathop{\operatorname{ess}\inf}_{Q \in \mathcal{D}_{t}}\alpha_{t,\lambda}(Q)\in L_{t}^{\infty}(P)$ for every $\lambda\in\Lambda$, 
such that for every $X \in L_{T}^{\infty}(P)$, 
$$ \rho_{t}(X)=\mathop{\operatorname{ess}\inf}_{\lambda \in \Lambda}\mathop{\operatorname{ess}\sup}_{Q \in \mathcal{D}_{t}}\Big(E_{Q}\left[-X \mid \mathscr {F}_{t}\right]-\alpha_{t,\lambda } (Q)\Big),$$ and the essential infimum is attained. Here, conditional convexity of penalties is understood through the normalized density mixtures.
		
(c) There exists a nonempty family $\left\{\rho_{t,\lambda} \mid \lambda \in \Lambda\right\}$ of continuous from above  dynamic convex risk measures on $L_{T}^{\infty}(P)$ such that $$\rho_{t}(X)=\mathop{\operatorname{ess}\inf}\limits_{\lambda \in \Lambda} \rho_{t,\lambda}(X)  ,\quad \forall \text X \in L_{T}^{\infty}(P),$$ and the essential infimum is attained.  Here,  continuity from above means that  $ \mathop{\lim}\limits_{n \rightarrow \infty} \rho_{t}\left(X_{n}\right)=\rho_{t}(X) $  for any sequence  $ \{X_{n}\}_{n\geq1} $  in $  L_T^{\infty}(P)$  decreasing to some $ X \in L_T^{\infty}(P)$. 
		
(d) For each $ X \in L_{T}^{\infty}(P)$, $$\rho_{t}(X)=\operatorname{ess}\inf \{h_{t}(X) \mid h_{t} \text{ is a dynamic convex risk measure and }  h_{t} \geq \rho_{t}\}.$$
\end{proposition}
		
\begin{proof} We only need to prove $ (a) \Rightarrow (b) $ and $ (c)\Rightarrow (d) $, since $ (b) \Rightarrow (c)$  and  $(d)\Rightarrow (a)$ are obvious.
	
$(a) \Rightarrow (b)$:  Suppose $ \rho_{t} $ is a dynamic monetary risk measure.  For any  fixed $Z \in \mathscr{A}_{t}:=\mathscr{A}_{\rho_t}$,  let
\begin{align}\label{atz1}
\mathscr{A}_{t}(Z):=Z+L_{T,+}^{\infty}(P).\end{align}
Then, one can see that $\mathscr{A}_{t}(Z)$ is nonempty, solid, conditionally convex and conditionally decomposable. Since 
$\operatorname{ess} \inf (L_{t}^{\infty}(P)\cap \mathscr{A}_{t}(Z))$ is bounded by $-||Z||_{\infty}$ and  $||Z||_{\infty}$, then (B2) holds. Besides, it is also conditionally-norm closed. Indeed, if $Y_{n}\geq Z$ and $||Y_{n}-Y||_{t}\rightarrow 0$, then $Y\geq Z-||Y_{n}-Y||_{t}$, and hence $Y\geq Z$. 
By  Lemma \ref{lem:5-1}, it follows that  
$\rho_{\mathscr{A}_{t}(Z)}(\cdot) $ is a dynamic convex risk measure and  $\mathscr{A}_{\rho_{\mathscr{A}_{t}(Z)}}=\mathscr{A}_{t}(Z)$.

To establish weak-star closedness of $\mathscr{A}_{t}(Z)$, observe that
$$\mathscr A_t(Z)=\bigcap_{H\in L_{T,+}^1(P)}
\{Y\in L_T^\infty(P)~~|~~E_P[(Y-Z)H]\geq0\}.$$
The forward inclusion follows from $Y\in\mathscr{A}_{t}(Z)$, or equivalently, $Y\geq Z$.  For the reverse inclusion, if $P(Y<Z)>0$, choosing $H=1_{\{Y<Z\}}$, it implies that $E_P[(Y-Z)H]<0$, a contradiction. 
Moreover, each set in the intersection is a weak-star closed half-space, because $Y\mapsto E_P[YH]$ is continuous for $\sigma(L_T^\infty(P),L_T^1(P))$.  Therefore, the intersection is weak-star closed.  

By choosing  the equivalent probability $\widetilde Q=P$, then 
$\inf_{Y\in\mathscr A_t(Z)}E_P[Y]=E_P[Z]>-\infty$. Hence, using  
the argument (IV) $\Rightarrow$ (I) of Theorem 3.1 in \cite{KS2007},  together with the weak-star closedness proved above,  it implies  the continuity from above of $\rho_{\mathscr{A}_{t}(Z)}(\cdot)$.  

Now we can apply Theorem 1 in \cite{DS2005}, with  conditioning sigma-field $\mathscr G=\mathscr F_t$ and the probability class $\mathcal P_{\mathscr G}=\mathcal D_t$.  Although the theorem is presented for a normalized dynamic convex risk measure, we can apply it to $\rho_{\mathscr A_t(Z)}-\rho_{\mathscr A_t(Z)}(0)$, then adding $\rho_{\mathscr A_t(Z)}(0)$ back, consistently with their Remark 2.   The resulting penalty for $\rho_{\mathscr A_t(Z)}$ is the minimal penalty for  the normalized  minus $\rho_{\mathscr A_t(Z)}(0)$.  Therefore, we obtain
$$
\rho_{\mathscr A_t(Z)}(X)=\operatorname{ess\,sup}_{Q\in\mathcal D_t}
\{E_Q[-X\mid\mathscr F_t]-\alpha_{t,Z}(Q)\},
$$
with the minimal penalty given by the acceptance-set formula in their Remark 9, i.e.,
$\alpha_{t,Z}(Q)=\operatorname{ess\,sup}_{Y\in\mathscr A_t(Z)}
E_Q[-Y\mid\mathscr F_t]
=E_Q[-Z\mid\mathscr F_t]
$.   Besides,  $|\alpha_{t,Z}(Q)|\leq\|Z\|_\infty$, so its essential infimum is bounded as required in (b).

For $L_1,L_2\in\mathcal L_t$ and $\beta\in L_t^\infty(P)$, $0\leq\beta\leq1$, the density mixture $L_\beta=\beta L_1+(1-\beta)L_2$ belongs to $\mathcal L_t$. Denote $Q^L$ for the probability with density $L$, and $\widetilde\alpha_{t,Z}(L)=\alpha_{t,Z}(Q^L)$. Then, 
\begin{align*}
\widetilde\alpha_{t,Z}(L_\beta)=-E_P[(\beta L_1+(1-\beta)L_2)Z\mid\mathscr F_t]=\beta\widetilde\alpha_{t,Z}(L_1)
 +(1-\beta)\widetilde\alpha_{t,Z}(L_2).
\end{align*}Hence, the constructed penalty is conditionally affine, and in particular conditionally convex.  Throughout, conditional convexity of penalties is understood with respect to these density mixtures, with the convention $0\cdot(+\infty)=0$ for extended values.

Noting that $\mathscr A_t(Z)\subseteq\mathscr A_t$ implies 
$\rho_{\mathscr A_t(Z)}\ge\rho_t$. 	
Moreover, for each $ X \in L_{T}^{\infty}(P)$,  we have $ Z_0:=X+\rho_{t}(X)  \in \mathscr{A}_{t}$,  and 
\begin{align*}
\rho_{\mathscr{A}_{t}(Z_0)}(X)&=\operatorname{ess}\inf\left\{Y\in L_{t}^{\infty}(P)\mid X+Y\in\mathscr{A}_{t}(Z_0)\right\}\nonumber\\
			&=\operatorname{ess}\inf\left\{Y\in L_{t}^{\infty}(P)\mid X+Y \geq X+\rho_{t}(X)\right\}=\rho_{t}(X).
\end{align*}Hence, we have that \begin{equation}\label{equa7}
\rho_{t}(X)=\rho_{\mathscr{A}_{t}}(X)=\mathop{\operatorname{ess}\inf}_{Z\in \mathscr{A}_{t}} \rho_{\mathscr{A}_{t}(Z) }(X) ,\quad \forall  X \in L_{T}^{\infty}(P),\end{equation}and the essential infimum in \eqref{equa7} is attainable.  Taking $\Lambda=\mathscr A_t$, it proves (b).

%Since $\mathscr{A}_{t}=\bigcup_{Z \in \mathscr{A}_{t}} \mathscr{A}_{t}(Z)$,   a lower-envelope representation result analogous to Lemma 2.6 in \cite{JXZ2020} yields \begin{equation}\label{equa7} \rho_{t}(X)=\rho_{\mathscr{A}_{t}}(X)=\mathop{\operatorname{ess}\inf}_{Z\in \mathscr{A}_{t}} \rho_{\mathscr{A}_{t}(Z) }(X) ,\quad \forall  X \in L_{T}^{\infty}(P).\end{equation} Moreover, for each $ X \in L_{T}^{\infty}(P)$,  we have $ Z_0:=X+\rho_{t}(X)  \in \mathscr{A}_{t}$,  and \begin{align*}\rho_{\mathscr{A}_{t}(Z_0)}(X)&=\operatorname{ess}\inf\left\{Y\in L_{t}^{\infty}(P)\mid X+Y\in\mathscr{A}_{t}(Z_0)\right\}\nonumber\\	&=\operatorname{ess}\inf\left\{Y\in L_{t}^{\infty}(P)\mid X+Y \geq X+\rho_{t}(X)\right\}=\rho_{t}(X),\end{align*}	

$(c) \Rightarrow (d)$:   $\rho_{t}(X)\leq\operatorname{ess}\inf \{h_{t}(X) \mid h_{t} $ is a dynamic convex risk measure and $h_{t}\geq\rho_{t}\}$  holds trivially for all $X\in L_{T}^{\infty}(P)$.   Conversely, for any $X\in L_T^\infty(P)$, attainment in (c) gives $\lambda^*\in\Lambda$ such that
$\rho_t(X)=\rho_{t,\lambda^*}(X)$. Since $\rho_{t,\lambda^*}$ is a dynamic convex risk measure dominating $\rho_t$, it belongs to the family in (d).
Thus the essential infimum in (d) is at most $\rho_t(X)$.
 \hfill$\Box$
\end{proof}

\begin{remark}	
Proposition \ref{prop:3-1} provides a dynamic counterpart of the static lower-envelope representation in \cite{JXZ2020}, Theorem 3.1.   Their part (b) uses countably additive probabilities, whereas part (c) obtains an inner maximum over the weak-star compact class of finitely additive probabilities. Here,  the representation over $\mathcal D_t$ follows from weak-star closedness of the constructed acceptance sets and the dynamic convex representation theorem.  The outer essential infimum is attained, while attainment of the inner essential supremum is not asserted.  The compactness argument used in their minimax proofs cannot simply be transferred to this setting. %This is a representation at each fixed time $t$. It does not, by itself, establish time consistency of the original risk measure or of the representing family. 
\end{remark}

\begin{remark}
In Proposition~\ref{prop:3-1}, using $\mathcal D_t$ makes conditional
expectations affine in normalized densities without changing the
constructed representation. Indeed, for $Q\in\mathscr Q_t$ with
density $L=dQ/dP$, the normalized density
$d\widehat Q/dP
=L/E_P[L|\mathscr F_t]
$
implies $\widehat Q\in\mathcal D_t$ and satisfies
$E_{\widehat Q}[H|\mathscr F_t]
=E_Q[H|\mathscr F_t]$ for every bounded $H$.
The denominator is positive $P$-almost surely.
Since $\mathcal D_t\subset\mathscr Q_t$, the two classes yield
the same conditional evaluations, including the constructed
penalties $\alpha_{t,Z}(Q)=E_Q[-Z|\mathscr F_t]$.
\end{remark}

\subsection{Representations of normalized dynamic star-shaped risk measures}

This subsection establishes the representation theorem for normalized dynamic star-shaped risk measures.  Prior to presenting the main representation theorem, we first introduce a bounded Koml\'{o}s-type convergence result of Lemma  A1.1 in \cite{DS1994},   which will be employed in subsequent proofs.

\begin{lemma}(Koml\'{o}s Lemma) \label{lem:5-2}
Let $\{a_n\}_{n\geq1}$ be a sequence in $L_t^\infty(P)$
such that $0\leq a_n\leq1$.
Then there exist integers $N_k\geq k$ and deterministic coefficients
$c_{kj}\geq0$, $j=k,\cdots,N_k$, satisfying
$\sum_{j=k}^{N_k}c_{kj}=1$, such that
$$
\bar a_k:=\sum_{j=k}^{N_k}c_{kj}a_j
\longrightarrow a
\qquad P\text{-a.s.},
$$
for some $a\in L_t^\infty(P)$ with $0\leq a\leq1$.
Here,  $N_k$ is the finite upper index of the $k$th convex
combination, formed from the tail $(a_j)_{j\geq k}$.
\end{lemma}

\begin{remark}\label{rem:3.2...}
Indeed, Lemma  A1.1 in \cite{DS1994} applies to general nonnegative measurable functions $f_n$ and provides almost surely convergence of forward convex combinations $g_n\in\operatorname{conv}(f_n,f_{n+1},\cdots)$, with a possibly infinite limit.  Lemma \ref{lem:5-2}
 is its direct specialization to $f_n=a_n\in[0,1]$ on $(\Omega,\mathscr F_t,P)$.  The finite sums displayed above are the convex hull formulation. Each $\bar a_{k}$ is $\mathscr F_t$-measurable and takes values in $[0,1]$, so its  limit also belongs to $L_t^\infty(P)$ and remains in $[0,1]$.

%The limit $g$ is automatically measurable as it is the almost sure limit of the measurable functions $g_n$. Moreover, since each $g_n$ is a convex combination of the original sequence, any pointwise bound on the $f_n$ (e.g., $f_n\in[0,1]$) is inherited by $g_n$ and, by passing to the limit, also by $g$. This stability property ensures that when the lemma is applied in the proof of Proposition~\ref{prop:3-2}, the limiting function $g$ belongs to $L_t^\infty(P)$ and takes values in $[0,1]$, which is essential for the closedness argument.
\end{remark}

\begin{proposition}\label{prop:3-2}
For a mapping $\rho_{t}:L_{T}^{\infty}(P)\rightarrow L_{t}^{\infty}(P)$, $t\in[0,T]$,  the following assertions are equivalent.
		
(a) $ \rho_{t} $ is a normalized dynamic star-shaped risk measure.
		
(b) There exists a nonempty family $\left\{\alpha_{t,\lambda} \mid \lambda \in \Lambda\right\}$ of conditionally convex functionals ${\alpha_{t,\lambda}}:\mathcal{D}_{t}\rightarrow L^{0}_{t}((-\infty, \infty])$  satisfying 
 $ \operatorname{ess\,inf}\limits_{Q \in \mathcal{D}_{t}}\alpha_{t,\lambda } (Q)=0 $ for every $ {\lambda \in \Lambda} $, such that $$ \rho_{t}(X)=\mathop{\operatorname{ess}\inf}_{\lambda\in\Lambda}\mathop{\operatorname{ess}\sup}_{Q \in \mathcal{D}_{t}}\Big(E_{Q}\left[-X \mid \mathscr {F}_{t}\right]-\alpha_{t,\lambda } (Q)\Big),\quad \forall \text X \in L_{T}^{\infty}(P),$$
 and the essential infimum  is attained.  Here, conditional convexity of penalties is also understood through the normalized density mixtures.% as used in Proposition \ref{prop:3-1}.
 	    
(c) There exists a nonempty family $\left\{\rho_{t,\lambda} \mid \lambda \in \Lambda\right\}$ of continuous from above, normalized dynamic convex risk measures on $L_{T}^{\infty}(P)$ such that
		$$\rho_{t}(X)=\mathop{\operatorname{ess}\inf} _{\lambda \in \Lambda} \rho_{t,\lambda}(X),\quad \forall\text X \in L_{T}^{\infty}(P),$$ and the essential infimum  is attained.
		
(d) For each $ X \in L_{T}^{\infty}(P) $,
		$$\rho_{t}(X)=\operatorname{ess}\inf \{h_{t}(X) \mid h_{t} \text{ is a normalized dynamic convex risk measure and }  h_{t} \geq \rho_{t}\}.$$	
\end{proposition}
\begin{proof}  As in Proposition  \ref{prop:3-1}, $(b)$ implies $(c)$ since the inner dual components are conditionally convex monetary risk measures continuous from above, and their values at zero are $-\operatorname{ess\,inf}_{Q\in\mathcal D_t}\alpha_{t,\lambda}(Q)=0$. 
$(c)\Rightarrow  (d)$ holds trivially.  In the following,  we prove $(a)$ implies $(b)$, and then verify $(d)$ implies $(a)$.

$(a)\Rightarrow(b)$. Suppose $ \rho_{t} $ is a normalized  dynamic  star-shaped risk measure. For any fixed $Z \in \mathscr{A}_{t}:=\mathscr{A}_{\rho_t}$, define 
\begin{align}\label{atz2}\mathscr{A}_{t}(Z):=\left\{X\in L_{T}^{\infty}(P)\mid X \geq \alpha Z \text { for some } \alpha \in L_{t}^{\infty}(P) \text{ and } \alpha \in[0,1] \right\} .\end{align}
Then each set $\mathscr{A}_{t}(Z)$ contains zero, and  is  solid and conditionally convex.  Besides, the conditional convexity also implies  (B7), by taking indicator weights. Noting $\mathscr{A}_{t}$ is conditionally star-shaped at zero and solid, which implies $\mathscr{A}_{t}(Z)\subset \mathscr{A}_{t}$.  For each  $m\in\mathscr A_t(Z)\cap L_t^\infty(P)$, then $0\geq\rho_t(m)=-m$, together with $0\in\mathscr A_t(Z)$, we can see that $\operatorname{ess} \inf (L_{t}^{\infty}(P)\cap \mathscr{A}_{t}(Z))=0$, i.e., (B2) and (B3) hold.

Next, we verify the  closedness of $\mathscr A_t(Z)$.  Suppose $Y_n\in\mathscr A_t(Z)$ and $Y_n\to Y$ almost surely, with $Y\in L_T^\infty(P)$.   Choose $0\leq a_n\leq1$ in $L_t^\infty(P)$ such that $Y_n\geq a_nZ$. By Lemma \ref{lem:5-2}, one can obtain that  
the forward convex combinations $\bar a_k=\sum_{j=k}^{N_k}c_{kj}a_j\to a$ almost surely, and $a\in L_t^\infty(P)$ with $0\leq a\leq1$.  
We use exactly the same coefficients for the positions, and derive that 
$$
\bar Y_k=\sum_{j=k}^{N_k}c_{kj}Y_j\geq\bar a_kZ,
\qquad
|\bar Y_k-Y|\leq\sup_{j\geq k}|Y_j-Y|\longrightarrow0
\quad P\text{-a.s.}.
$$
Therefore,  $Y\geq aZ$, i.e., $Y\in\mathscr A_t(Z)$.   In particular, $\mathscr A_t(Z)$ is also conditionally-norm closed.
Indeed, if $Y_n\in\mathscr A_t(Z)$ and
$\|Y_n-Y\|_t\to0$ almost surely, then
$|Y_n-Y|\leq\|Y_n-Y\|_t$ implies $Y_n\to Y$ almost surely.

Furthermore, the almost surely closedness established above also implies that,
for every $\gamma>0$, 
$\mathscr A_t(Z)\cap
\{Y\in L_T^\infty(P):\|Y\|_\infty\leq\gamma\}
$ is closed in $L_T^1(P)$.  Since $\mathscr A_t(Z)$ is convex, Lemma A.68 of \cite{FS16} implies that it is weak-star closed in
$\sigma(L_T^\infty(P),L_T^1(P))$.

From Lemma  \ref{lem:5-1}, we know that $\rho_{\mathscr A_t(Z)}$ is a  normalized dynamic convex risk measure.  Noting that $\inf_{Y\in\mathscr A_t(Z)}E_P[Y]\geq-\|Z\|_\infty>-\infty$, then by Theorem 3.1 in  \cite{KS2007} as in Proposition \ref{prop:3-1}, one can get $\rho_{\mathscr A_t(Z)}$ is continuous from above.  Thus, by Theorem 1 and Remark 9 of \cite{DS2005}, we derive that 
$$\rho_{\mathscr A_t(Z)}(X)
=\operatorname{ess\,sup}_{Q\in\mathcal D_t}
\big\{E_Q[-X\mid\mathscr F_t]-\alpha_{t,Z}(Q)\big\},\quad \forall \text X \in L_{T}^{\infty}(P),$$with 
$\alpha_{t,Z}(Q)
=\operatorname{ess\,sup}_{Y\in\mathscr A_t(Z)}E_Q[-Y\mid\mathscr F_t]=\big(E_Q[-Z\mid\mathscr F_t]\big)^+$. For the equality of $\alpha_{t,Z}(Q)$, indeed, for  $Y\geq aZ$, it implies that $E_Q[-Y\mid\mathscr F_t]\leq aE_Q[-Z\mid\mathscr F_t]\leq(E_Q[-Z\mid\mathscr F_t])^+$, and  the upper bound is attained by $Y=aZ$ with $a=1_{\{E_Q[-Z\mid\mathscr F_t]\geq0\}}$.

We verify the conditional convexity of penalties $\alpha_{t,Z}(\cdot)$. In fact, we denote $\widetilde\alpha_{t,Z}(L)=\alpha_{t,Z}(Q^L)=(E_P[-LZ\mid\mathscr F_t])^+$. Obviously, by conditional linearity and convexity of the  function $x^{+}$,  it is not hard to prove that for $\beta\in L_t^\infty(P)$ with $0\leq\beta\leq1$, 
$$\widetilde\alpha_{t,Z}(L_\beta)
\leq\beta\widetilde\alpha_{t,Z}(L_1)
+(1-\beta)\widetilde\alpha_{t,Z}(L_2), \quad L_1,L_2\in\mathcal L_t, $$
where  $L_\beta=\beta L_1+(1-\beta)L_2$.  Besides, $\operatorname{ess\,inf}_{Q\in\mathcal D_t}\alpha_{t,Z}(Q)
=-\rho_{\mathscr A_t(Z)}(0)=0.$

Since the acceptance set $\mathscr A_t(Z)\subset \mathscr A_t$, then   $\rho_{\mathscr A_t(Z)}\geq\rho_t$.   For any given $X\in L_T^\infty(P)$, let $Z_0=X+\rho_t(X)\in\mathscr A_t$. Taking $a=1$,   then $Z_0\in\mathscr A_t(Z_0)$, and thereby $\rho_{\mathscr A_t(Z_0)}(X)\leq\rho_t(X)$. 
 Consequently,
$$\rho_t(X)=\rho_{\mathscr A_t}(X)
=\operatorname{ess\,inf}_{Z\in\mathscr A_t}
\rho_{\mathscr A_t(Z)}(X),
\qquad X\in L_T^\infty(P),
$$
with the essential infimum attained at $Z_0$.  We complete the proof of $(b)$.

 $(d)\Rightarrow(a)$.  Monotonicity, conditional translation invariance and normalization follow immediately.    For any normalized dynamic convex risk measure such that  $h_t\geq \rho_{t}$ and $a\in L_t^\infty(P)$ with $a\geq1$, then the conditional convexity of  $h_t$ implies
$
h_t(aX)\geq a h_t(X)\geq a\rho_t(X).
$
Taking the essential infimum over $h_{t}$ in $(d)$, one gets the conditional star-shapedness of $\rho_t$.  \hfill$\Box$
\end{proof}

\begin{remark}
Proposition \ref{prop:3-2}  yields a conclusion analogous to that of  Proposition \ref{prop:3-1}. In Proposition \ref{prop:3-1}, we show that a dynamic monetary risk measure is the lower envelope of a family of dynamic convex risk measures, while by contrast, a normalized dynamic star-shaped risk measure is the lower envelope of a family of \textit{normalized} dynamic convex risk measures.
In subsection 3.3,  we will elaborate on the intrinsic link  between dynamic monetary risk measures and dynamic star-shaped risk measures. %It turns out that, under mild conditions, a dynamic monetary risk measure is merely a translation of a dynamic star-shaped risk measure.
\end{remark}

A number of dynamic risk measures are star-shaped,  for instance, dynamic normalized convex risk measures, dynamic VaR, among others.  Here we provide a specific example of utility-based risk measures.

\begin{example}\label{starlizi}(Star-shaped utilities) Let $u:\mathbb{R}\rightarrow \mathbb{R} $ be a continuous and strictly 
increasing function with $u(0)=0$. We also assume that $u(\lambda x)\leq \lambda u(x)$ for $x\in\mathbb{R}$ and $\lambda\geq1$. 
Define the following utility-based  risk measure
\begin{align}\label{utility}
\rho_{t}(X):=\mathop{\operatorname{ess}}\inf \{ m \in L^{\infty}_{t}(P)~| ~E_{P}[ u(X+m) ~| \mathscr{F}_{t}]\geq 0 \}, ~~~~X\in L^{\infty}_{T}(P).\end{align}Monotonicity and conditional translation invariance directly follow from the definition. We can easily verify that $-||X||_{\infty}\leq \rho_{t}(X)\leq ||X||_{\infty}$. When $X=0$, $u(m)\geq0$ is equivalent to $m\geq0$ by strict increase and $u(0)=0$. Hence, $\rho_t(0)=0$. Therefore, $\rho_t$ is a normalized dynamic monetary risk measure. 

To prove conditional star-shapedness,  take $a\in L_t^\infty(P)$ with $a\geq1$, and let $m\in L_t^\infty(P)$ be any admissible capital amount for $aX$. Then, by the property of $u$,  we get 
\begin{align*}
0\leq E_P[u(aX+m)| \mathscr F_t]\leq a E_P\left[u\left(X+\frac {m}{a}\right) | \mathscr F_t\right],
\end{align*}which means $m/a$ is acceptable for $X$, i.e., $\rho_{t}(X)\leq m/a$. Taking the essential infimum for all the above $m$, we can get $\rho_{t}(aX)\geq a\rho_{t}(X)$.

If $u(x)=x$, then $\rho_t(X)=-E_P[X|\mathscr F_t]$, which is 
consistent with the payoff convention. When  $u(x)=x^+$, it violates the strict monotonicity, and every $m\in L_{t}^{\infty}(P)$ would satisfy the constraint,  resulting in $\rho_t(X)=-\infty$. 

At $t=0$, this definition reduces to a static utility-based
shortfall risk measure.  Its star-shapedness in the static setting is discussed in \cite{CCMW2022}; see also \cite{FS16} for the utility-based shortfall framework. Moreover, $\rho_t$ is conditionally law-invariant, since the admissibility condition depends only on the conditional distribution of $X$ given $\mathscr F_t$. We refer to \cite{LRZ2024} for the related static results on law-invariant star-shaped risk measures.\end{example}

The following corollary  establishes  that a dynamic positively homogeneous risk measure is the lower envelope of a family of dynamic coherent risk measures. 
	
\begin{corollary}\label{cor:3-1}
For a mapping $\rho_{t}:L_{T}^{\infty}(P)\rightarrow L_{t}^{\infty}(P)$, $t\in[0,T]$, then the following assertions are equivalent.
		
(a) $ \rho_{t} $ is a dynamic positively homogeneous risk measure.
		
%(b) There exists a family $\left\{\mathcal{P}_{\lambda} \mid \lambda \in \Lambda\right\}$ of convex subsets on $ \mathscr{Q}_{t}$ such that $$\rho_{t}(X)=\mathop{\operatorname{ess}\inf}_{\lambda \in \Lambda}\mathop{\operatorname{ess}\sup}_{Q \in \mathcal{P}_{\lambda}}E_{Q}\left[-X \mid \mathscr{F}_{t}\right],\quad \forall\text X \in L_{T}^{\infty}(P),$$and the essential infimum  is attained.
		
(b) There exists a nonempty family $\left\{\rho_{t, \lambda} | \lambda \in \Lambda\right\}$ of  dynamic coherent risk measures on $L_{T}^{\infty}(P)$ such that
$$\rho_{t}(X)=\mathop{\operatorname{ess}\inf} _{\lambda \in \Lambda} \rho_{t,\lambda}(X),\quad \forall\text X \in L_{T}^{\infty}(P),$$and the essential infimum is  attained.
		
(c) For each $ X \in L_{T}^{\infty}(P) $,
$$\rho_{t}(X)=\operatorname{ess}\inf \{h_{t}(X) \mid h_{t} \text{ is a dynamic coherent risk measure and }  h_{t} \geq \rho_{t}\} .$$	
\end{corollary}

\begin{proof}  $(a) \Rightarrow (b)$. Suppose $ \rho_{t} $ is a dynamic positively homogeneous risk measure.  For any $ Z \in \mathscr{A}_{t}:=\mathscr{A}_{\rho_t}$,  define
\begin{align}\label{eq:posi}
\mathscr{A}_{t}(Z):=\left\{X\in L_{T}^{\infty}(P)\mid X \geq \alpha Z \text { for some } \alpha \in L_{t,+}^{\infty}(P)\right\}.
\end{align}
We can see that $\mathscr{A}_{t}(Z)$ contains zero and $Z$, and is solid, conditionally convex, conditionally conic and conditionally decomposable.   For any $a\in L^{\infty}_{t,+}(P)$,  $\rho_t(aZ)=a\rho_t(Z)\leq0$, which implies $\mathscr A_t(Z)\subset\mathscr A_t$. For any $m\in\mathscr A_t(Z)\cap L_t^\infty(P)\subset \mathscr A_t$, it means that $0\geq\rho_t(m)=-m$. Hence,  $\operatorname{ess} \inf (\mathscr A_t(Z)\cap L_t^\infty(P))=0$.

By the construction part of Lemma \ref{lem:5-1}, the map
$$
\rho_{\mathscr A_t(Z)}(X)
=\operatorname{ess\,inf}\{m\in L_t^\infty(P) ~|~ X+m\in\mathscr A_t(Z)\}
$$
is a dynamic coherent risk measure. This construction does not require closedness. In Lemma \ref{lem:5-1}, conditional-norm closedness is used in the acceptance set reconstruction, which is not needed here.

Since  $\mathscr A_t(Z)\subset \mathscr A_t$,  then $\rho_{\mathscr A_t(Z)}\geq\rho_t$.   For any given $X\in L_T^\infty(P)$, let $Z_0=X+\rho_t(X)\in\mathscr A_t$. Taking $a=1$,   then $Z_0\in\mathscr A_t(Z_0)$, and thereby $\rho_{\mathscr A_t(Z_0)}(X)\leq\rho_t(X)$. 
Therefore,
$$\rho_t(X)=\rho_{\mathscr A_t}(X)
=\operatorname{ess\,inf}_{Z\in\mathscr A_t}
\rho_{\mathscr A_t(Z)}(X),
\qquad X\in L_T^\infty(P),
$$with the essential infimum attained at $Z_0$.  It completes the proof of $(b)$.

The part $(b)\Rightarrow(c)$ is obvious.  For $(c)\Rightarrow(a)$, we only verify the positive homogeneity.   For $k\in L_t^\infty(P)$ and $k\geq\epsilon>0$, then $\rho_t(kX)=\operatorname{ess\,inf}_{h_t\geq\rho_t}k h_t(X)=k\rho_t(X)$.  For a general bounded $k\geq0$, replace $k$ by $k+1/n$ and pass to the limit using the Lipschitz continuity of monetary risk measures.
\hfill$\Box$
\end{proof}

\begin{remark}
Unlike Propositions \ref{prop:3-1} and \ref{prop:3-2}, Corollary \ref{cor:3-1} does not ensure continuity from above or a countably additive dual representation for the components.  Lemma \ref{lem:5-1} constructs $\rho_{\mathscr A_t(Z)}$ without requiring closedness of $\mathscr A_t(Z)$. The actual acceptance sets
$\mathscr A_{\rho_{\mathscr A_t(Z)}}$ are conditionally-norm closed, but need not be weak-star closed or coincide with $\mathscr A_t(Z)$.

We thank the referee for providing the following counterexample. Consider the static case on $\Omega=(0,1)$ with Lebesgue probability $P$.  For every $X\in L_{T}^{\infty}(P)$, define $\rho(X)=-\mathop{\operatorname{ess}\sup} X$. Then it is a normalized monetary, positively homogeneous risk measure. For $U(\omega)=\omega$, we have 
$\rho(-U)=0<\sup_{Q\in\mathcal{Q}}E_Q[U],$ where $\emptyset\neq\mathcal{Q}\subset\mathscr{M}_{1}(P)$. Hence,  no coherent component represented by such probabilities can attain $\rho(-U)$. Since coherent risk measures continuous from above admit such representations, an attained envelope of such components need not exist. 
Furthermore, the cone $\mathscr{A}_{0}(Z)$ defined by \eqref{eq:posi}
need not be norm closed. Indeed, for $Z=-U$, the sequence $Y_n=-\min\{\sqrt U,nU\}$ belongs to $\mathscr A_0(Z)$ and satisfies
$
\|Y_n+\sqrt U\|_\infty=\frac{1}{4n}\longrightarrow0.
$
However, $-\sqrt U\notin\mathscr A_0(Z)$, since it would require $U^{-1/2}\leq a$  for some finite constant $a$.  
\end{remark}

\begin{remark}
Corollary \ref{cor:3-1} represents dynamic positively homogeneous risk measures as attained lower envelopes of dynamic coherent risk measures. This parallels the representation in \cite{C2000} of proper positively homogeneous functions vanishing at the origin as pointwise minima of sublinear majorants. The conditional formulation uses essential infima and does not imply a countably additive probability representation.
\end{remark}

\begin{example} (Conditional VaR)  For any $\lambda\in(0,1)$, define
$$\mathscr A_t=\{X\in L_T^\infty(P):P(X<0\mid\mathscr F_t)\leq\lambda\}.$$
The corresponding capital requirement, called conditional Value at Risk at level $\lambda$, is
$$
\operatorname{VaR}_\lambda^P(X\mid\mathscr F_t)
:=\operatorname{ess\,inf}
\{m\in L_t^\infty(P):P(X+m<0\mid\mathscr F_t)\leq\lambda\}.
$$
It is a normalized dynamic positively homogeneous risk measure, with
$\bigl|\operatorname{VaR}_\lambda^P(X\mid\mathscr F_t)\bigr|\leq\|X\|_\infty$.
It is not conditionally convex in general. By Corollary \ref{cor:3-1},
$$
\operatorname{VaR}_\lambda^P(X\mid\mathscr F_t)
=\operatorname{ess\,inf}\left\{h_t(X):
\begin{array}{l}
h_t\text{ is a dynamic coherent risk measure},\\
h_t(\cdot)\geq\operatorname{VaR}_\lambda^P(\cdot\mid\mathscr F_t)
\end{array}\right\},
$$
and the essential infimum is attained. Furthermore, for a nonempty collection $\mathscr P\subseteq\mathscr Q_T$, define the robust conditional VaR (or scenario-based VaR) by
$$
\operatorname{ess\,sup}_{Q\in\mathscr P}
\operatorname{VaR}_\lambda^Q(X\mid\mathscr F_t),
$$
where $\operatorname{VaR}_\lambda^Q$ is defined under $Q$. This is also a normalized dynamic positively homogeneous risk measure.  
For the representation of positively homogeneous functions by
sublinear majorants, see \cite{C2000}; for further discussion of
scenario-based risk measures, see \cite{WZ2021,NPS2008}.
\end{example}

\subsection {The link between dynamic monetary and dynamic star-shaped risk measures}

\cite{MR2022}  established a subtle relationship between the results of \cite{JXZ2020} and \cite{CCMW2022}. In this subsection, we establish an analogous dynamic version of the relationship between dynamic monetary risk measures and dynamic star-shaped risk measures.
Before proceeding, we first present a preliminary proposition that will be employed in the subsequent analysis.

\begin{proposition}\label{prop:3-4}
Let $\{\left(\rho_{t,\lambda}\right)_{0 \leq t \leq T} \mid \lambda \in \Lambda\}$ be a nonempty family of dynamic monetary risk measures.  For each $t\in[0,T]$ and $\lambda\in\Lambda$, let $\mathscr{A}_{t,\lambda}$ denote the acceptance set of $\rho_{t,\lambda}$, and define $\mathscr{B}_{t}=  \cap_{\lambda\in\Lambda} \mathscr{A}_{t,\lambda}$.   Then, we have 
$$\mathscr{B}_t \cap L_{t}^{\infty}(P) \neq \emptyset \Longleftrightarrow \mathscr{B}_t \neq \emptyset \Longleftrightarrow \operatorname{ess}\sup _{\lambda\in\Lambda} \rho_{t,\lambda}(0)\in L_{t}^{\infty}(P). $$
\end{proposition} 

\begin{proof}  For any $t\in[0,T]$ and $\lambda\in\Lambda$,  since $ \mathscr{A}_{t,\lambda}$ is the acceptance set  of dynamic monetary risk measure $\rho_{t,\lambda}(\cdot)$,  it implies $ \mathscr{A}_{t,\lambda}$ is solid.   This
in turn implies the solidity of $ \mathscr{B}_t$.  
If $Y\in \mathscr{B}_t$, then the finite constant $C=\operatorname{ess}\sup Y$ also belongs to $\mathscr{B}_t$.
Hence, $\mathscr{B}_t \cap L_{t}^{\infty}(P) \neq \emptyset \Longleftrightarrow \mathscr{B}_t \neq \emptyset$.

We claim that if $\mathscr{B}_t \neq \emptyset$, then
\begin{align}  \label{denghao}
\rho_{\mathscr{B}_t}(X)=\operatorname{ess}\sup _{\lambda\in\Lambda} \rho_{t,\lambda}(X),  \quad\forall X\in  L_T^{\infty}(P). \end{align} 
In fact, fix $\lambda_{0} \in \Lambda$ and  $Y\in\mathscr{B}_t \subset \mathscr{A}_{t, \lambda_{0}}$.  Then for any $X\in L_T^{\infty}(P)$, we have
\begin{align*}
\rho_{\mathscr{B}_t}(X) =\operatorname{ess} \inf \left\{m \in L_t^{\infty}(P) \mid X+m \in \mathscr{B}_t\right\} \leq  ||X||_{\infty}  + \operatorname{ess} \sup Y<+\infty,
\end{align*}and 
$\rho_{{t,\lambda_{0}}}(X)\leq \operatorname{ess}\sup _{\lambda\in\Lambda} \rho_{t,\lambda}(X)\leq \rho_{\mathscr{B}_t}(X).$
Thus, $\rho_{\mathscr{B}_t}$ and $ \operatorname{ess}\sup _{\lambda\in\Lambda} \rho_{t,\lambda}$ are both valued in $L_t^\infty(P)$.

Moreover, for any $X\in L_T^{\infty}(P)$,  conditional translation invariance yields 
\begin{align*}
\operatorname{ess\,sup}_{\lambda\in\Lambda}\rho_{t,\lambda}(X)
&=\operatorname{ess\,inf}\left\{m\in L_t^\infty(P) ~|~
\operatorname{ess\,sup}_{\lambda\in\Lambda}\rho_{t,\lambda}(X)\leq m\right\}\\
&=\operatorname{ess\,inf}\{m\in L_t^\infty(P) ~|~
\rho_{t,\lambda}(X)\leq m\text{ for every }\lambda\in\Lambda\}\\
&=\operatorname{ess\,inf}\{m\in L_t^\infty(P) ~|~
X+m\in\mathscr A_{t,\lambda}\text{ for every }\lambda\in\Lambda\}\\
&=\operatorname{ess\,inf}\{m\in L_t^\infty(P) ~|~X+m\in\mathscr B_t\}\\
&=\rho_{\mathscr B_t}(X).
\end{align*}
In particular, taking $X=0$, then $\operatorname{ess}\sup _{\lambda\in\Lambda} \rho_{t,\lambda}(0)= \rho_{\mathscr{B}_t}(0)\in L_{t}^{\infty}(P).$

Conversely, suppose $\operatorname{ess}\sup _{\lambda\in\Lambda} \rho_{t,\lambda}(0)\in L_{t}^{\infty}(P)$. 
Then, for every $\lambda\in\Lambda$, 
$$\rho_{t,\lambda}\big(\operatorname{ess}\sup _{\lambda\in\Lambda} \rho_{t,\lambda}(0)\big)=\rho_{t,\lambda}(0)-\operatorname{ess}\sup _{\lambda\in\Lambda} \rho_{t,\lambda}(0)\leq 0,$$which means 
$\operatorname{ess}\sup _{\lambda\in\Lambda} \rho_{t,\lambda}(0)\in \mathscr{A}_{t,\lambda}$.  Hence, $\operatorname{ess}\sup _{\lambda\in\Lambda} \rho_{t,\lambda}(0)\in\mathscr{B}_t \neq \emptyset$. 
\hfill$\Box$
\end{proof}	

\begin{remark}
We thank the referee for the following example, 
which shows that almost-sure finiteness of $\operatorname{ess}\sup _{\lambda\in\Lambda} \rho_{t,\lambda}(0)$ alone is insufficient.
Let $Y\geq0$ be $\mathscr F_t$-measurable, finite almost surely
but essentially unbounded, and define
$$
\rho_{t,n}(X)=(Y\wedge n)-E_P[X\mid\mathscr F_t],
\quad n\geq1,  \quad X\in L_T^\infty(P).
$$
Then each $\rho_{t,n}$ is a monetary risk measure with values in
$L_t^\infty(P)$, but $\sup_n\rho_{t,n}(0)=Y<+\infty$
almost surely.  A common accepted position $X\in L_T^\infty(P)$
would satisfy  $\rho_{t,n}(X)\leq0$ for all $n\geq1$, which means 
$E_P[X\mid\mathscr F_t]\geq Y\wedge n$
for every $n$, hence $E_P[X\mid\mathscr F_t]\geq Y$.
This is impossible because the conditional expectation is
essentially bounded.  Hence,  $\mathscr{B}_{t}=  \cap_{n\geq1} \mathscr{A}_{t,n}$ is empty. 
\end{remark}

The following proposition shows  that, under mild conditions, a dynamic monetary risk measure is merely a translation of a dynamic star-shaped risk measure.

%\red{the following to be revised}

\begin{proposition}\label{prop:3-5} 
	Let  $\{\rho_t \}_{t\in[0,T]}$  be a dynamic monetary risk measure. Fix  $t\in[0,T]$.  Then there  exists  $Z \in L_T^\infty(P)$ such that  $\rho_{t,Z}(X):=\rho_t(X+Z)$, $X \in L_{T}^{\infty}(P)$,   is a dynamic star-shaped monetary risk measure  if and only if $\rho_{t}$ admits a representation
\begin{align}\label{equa4}
\rho_{t}(X)=\mathop{\operatorname{ess}\inf}\limits_{\lambda \in \Lambda} \rho_{t,\lambda}(X), \quad \forall  X \in L_{T}^{\infty}(P),
		\end{align}by a nonempty family of dynamic convex risk measures satisfying  $\mathop{\operatorname{ess}\sup}_{\lambda\in\Lambda} \rho_{t,\lambda}(0)\in L_{t}^{\infty}(P)$.  In this case, one may  take any $ Z\in \cap_{\lambda\in\Lambda} \mathscr{A}_{\rho_{t,\lambda}}.$
\end{proposition}

\begin{proof}  ``$\Longleftarrow$''  Set $ \mathscr{B}_{t}:=  \cap_{\lambda\in\Lambda} \mathscr{A}_{t,\lambda}$  and  $\mathscr{A}_{t}:=  \cup_{\lambda\in\Lambda} \mathscr{A}_{t,\lambda}$, where $ \mathscr{A}_{t,\lambda}:=\mathscr{A}_{\rho_{t,\lambda}}$  is the  acceptance set of the dynamic convex risk measure $\rho_{t,\lambda} (\cdot)$ for each $\lambda\in \Lambda$.   For  $Y \in \mathscr{A}_{t}$,  choose  $\lambda\in\Lambda$ such that  $Y \in \mathscr{A}_{t,\lambda}$.  Since $ \mathop{\operatorname{ess}\sup}_{ \lambda\in\Lambda} \rho_{t,\lambda}(0)\in L_{t}^{\infty}(P) $,  Proposition \ref{prop:3-4}  implies  that $\mathscr{B}_t\neq \emptyset$.  
Thus,  for any $Z\in\mathscr{B}_{t} $ and $k \in L_{t}^{\infty}(P)$ with $0\leq k \leq 1$,  the conditional convexity of $\mathscr{A}_{t,\lambda}$ yields    $$k Y+(1-k)Z \in \mathscr{A}_{t,\lambda} \subset \mathscr{A}_{t},$$
which means that  $\mathscr{A}_{t}$ is conditionally star-shaped at $\mathscr{B}_{t}$. 
   
For any fixed $Z\in\mathscr{B}_{t}$, define
$$\mathscr{A}_{t}(Z):=\mathscr{A}_{t}-Z=\{Y-Z\mid Y \in \mathscr{A}_{t}\}.$$
Then $\mathscr{A}_{t}(Z)$ is non-empty and solid.  Moreover,  for any  $X \in \mathscr{A}_{t}(Z) $,  it follows that $  X+Z \in \mathscr{A}_{t}$.   The conditional star-shapedness of  $\mathscr{A}_{t} $  at $Z$  implies that  for any  $k \in L_{t}^{\infty}(P)$ with $0\leq k \leq 1$,  $$k(X+Z)+(1-k) Z=k X+Z \in \mathscr{A}_{t},$$which is equivalent to $k X \in \mathscr{A}_{t}(Z)$.   Hence,  $\mathscr{A}_{t}(Z)$ is  conditionally  star-shaped at $0$. 
   
 In addition, for any $X \in L_{T}^{\infty}(P)$, one obtains
\begin{align*}
\rho_{\mathscr{A}_{t}(Z)}(X) &=\operatorname{ess}\inf \{Y \in L_{t}^{\infty}(P) \mid X+Y \in \mathscr{A}_{t}(Z)\} \\
			&=\operatorname{ess}\inf \{Y \in L_{t}^{\infty}(P)\mid X+Y \in \mathscr{A}_{t}-Z\} \\
			&=\operatorname{ess}\inf \{Y \in L_{t}^{\infty}(P) \mid X+Z+Y \in \mathscr{A}_{t}\}\\
			&=\rho_{\mathscr{A}_{t}}(X+Z)\\
			&=\mathop{\operatorname{ess}\inf}\limits_{\lambda \in \Lambda} \rho_{t,\lambda}(X+Z) =\rho_{t}(X+Z)=:\rho_{t,Z}(X),
\end{align*}
where the antepenultimate equality follows from $\mathscr{A}_{t}:=  \cup_{\lambda\in\Lambda} \mathscr{A}_{t,\lambda}$.  
Here (B7) is not required for the generating set $\mathscr A_t(Z)$, and we verify (A6) directly. Let $a\in L_t^\infty(P)$, $a\geq1$. If $aX+m\in\mathscr A_t(Z)$, its conditional star-shapedness gives $X+m/a\in\mathscr A_t(Z)$, and hence $a\rho_{t,Z}(X)\leq m$. Taking the essential infimum over such $m$ proves
$\rho_{t,Z}(aX)\geq a\rho_{t,Z}(X).$
Monotonicity and conditional cash invariance follow from the definition of $\rho_{t,Z}$. Therefore, $ \rho_{t,Z}(\cdot)=\rho_{\mathscr{A}_{t}(Z)}(\cdot) $ is a dynamic star-shaped monetary risk measure.

``$\Longrightarrow$''  Suppose that for some $Z \in L_T^\infty(P)$,  the mapping  $\rho_{t,Z}(\cdot):=\rho_t(\cdot+Z) $ is a dynamic star-shaped monetary risk measure.    Since $ \rho_{t,Z}(\cdot) $  is conditionally star-shaped,  by Lemma \ref{lem:5-1}, it follows that  $ \mathscr{A}_{\rho_{t,Z}}=\mathscr{A}_{\rho_t}-Z$  is conditionally star-shaped at $0$.  Consequently,  $\mathscr{A}_{\rho_t}$ is conditionally star-shaped at $Z$.  Besides, $\rho_{t}(Z)=\rho_{t,Z}(0)\leq0$, and it implies that $Z\in \mathscr{A}_{\rho_t}$.

For any $ U\in \mathscr{A}_{\rho_t}$,   define
	$$
	\mathscr{B}_{t}(U):=\left\{V \in L_T^{\infty}(P)\mid V \geq U\right\} \text{~~and~~} \mathscr{B}_{Z}:=\left\{X \in L_T^{\infty}(P)\mid X \geq Z \right\}.
	$$		
Define their conditional convex hull, 
$$
\mathscr B_{t,U}
=\{V\in L_T^\infty(P)~|~
V\geq kU+(1-k)Z
\text{ for some }k\in L_t^\infty(P),\ 0\leq k\leq1\}.
$$Since $\mathscr A_{\rho_t}$ is conditionally star-shaped
at $Z$ and solid, we have
$$\mathscr B_t(U)\subseteq\mathscr B_{t,U}\subseteq\mathscr A_{\rho_t},
\qquad
\mathscr A_{\rho_t}=\bigcup_{U\in\mathscr A_{\rho_t}}\mathscr B_{t,U},
\qquad
Z\in\bigcap_{U\in\mathscr A_{\rho_t}}\mathscr B_{t,U}.$$
%Clearly,  $\mathscr{B}_{t}(U)$ and $\mathscr{B}_{Z}$ are both 	conditionally convex and solid, and $ \mathscr{A}_{\rho_t}=\bigcup_{U \in \mathscr{A}_{\rho_t}} \mathscr{B}_t(U)$.  Since $\mathscr{A}_{\rho_t}$ is conditionally star-shaped at $Z$,  together with the solidity of $ \mathscr{A}_{\rho_t} $,  we obtain $$\lambda\mathscr{B}_{t}(U)+(1-\lambda)\mathscr{B}_{Z}\subseteq\mathscr{A}_{\rho_t}.$$
%Therefore, for  every $ U\in \mathscr{A}_{\rho_t}$, we have  $ \mathscr{B}_{t,U} :=\operatorname{conv}\left(\mathscr{B}_{t}(U) \cup \mathscr{B}_{Z}\right) \subseteq\mathscr{A}_{\rho_t}$.  It follows that $$ \mathscr{A}_{\rho_t}=\bigcup_{U\in  \mathscr{A}_{\rho_t}} \mathscr{B}_{t,U} ~~\text{and}~~~~Z\in \bigcap_{U\in  \mathscr{A}_{\rho_t}} \mathscr{B}_{t,U}\neq \emptyset. $$  
Each $\mathscr B_{t,U}$ is nonempty, solid and conditionally convex, hence conditionally decomposable. Since  $\mathscr B_t(U)\subseteq\mathscr B_{t,U}\subseteq\mathscr A_{\rho_t}$, it implies that $ \rho_t(0)\leq \operatorname{ess\,inf}
\bigl(\mathscr B_{t,U}\cap L_t^\infty(P)\bigr)\leq ||Z||_{\infty}$.  From Lemma \ref{lem:5-1}, it induces a dynamic convex risk measure $\rho_{\mathscr B_{t,U}}$ with $\rho_{\mathscr B_{t,U}}\geq\rho_t$.
Since $Z\in\bigcap_{U\in\mathscr A_{\rho_t}}\mathscr B_{t,U}$, by Proposition \ref{prop:3-4},  we get  $ \mathop{\operatorname{ess}\sup}_{U\in  \mathscr{A}_{\rho_t}} \rho_{\mathscr{B}_{t,U} }(0)\in L_{t}^{\infty}(P)$.
	
Finally,  we show that the essential infimum is attained. 
For any $X\in L_T^\infty(P)$, set $U^*=X+\rho_t(X)\in\mathscr A_{\rho_t}$. Since $U^*\in\mathscr B_t(U^*)\subseteq\mathscr B_{t,U^*}$, then
$$
\rho_t(X)\leq\rho_{\mathscr B_{t,U^*}}(X)\leq\rho_t(X).
$$
Thus the essential infimum is attained at $U^*$. \hfill$\Box$
\end{proof}

\begin{remark}
Comparing Proposition \ref{prop:3-1} with Proposition \ref{prop:3-2}, one can see that  in both cases the risk measure is expressed as the essential infimum of a family of dynamic convex risk measures, albeit with different structural requirements  (normalization in the latter). Furthermore, Proposition \ref{prop:3-5} shows that dynamic monetary risk measures are  merely a translation of dynamic star-shaped risk measures whenever the family of dynamic convex risk measures satisfies $ \mathop{\operatorname{ess}\sup}_{\lambda\in\Lambda} \rho_{t,\lambda}(0)\in L_{t}^{\infty}(P)$, or equivalently, when the intersection of the corresponding convex acceptance sets is non-empty.

Proposition \ref{prop:3-5}  is a fixed-time result, so the translating position $Z$ may depend on $t$.  For a given representing family, any position in
$$
\bigcap_{t\in[0,T]}\bigcap_{\lambda\in\Lambda}
\mathscr A_{\rho_{t,\lambda}}
$$
provides a translation valid at all times. The separate conditions $\operatorname{ess\,sup}_{\lambda}\rho_{t,\lambda}(0)\in L_t^\infty(P)$ do not guarantee that this intersection is nonempty.  Moreover, $\rho_{t,Z}(0)=\rho_t(Z)\leq0$; the translated risk measure is normalized precisely when equality holds.
\end{remark}

\subsection{Sensitivity and time consistency}
In this subsection,  we discuss the sensitivity and time consistency of  dynamic monetary or star-shaped risk measures.
		
\begin{definition}\label{df:3.1.}
The dynamic monetary risk measure $\left\{\rho_{t}\right\}_{0 \leq t \leq T}$ is called relevant or sensitive if for every  $t\in[0,T]$,  there exists $ \tilde{Q} \in \mathcal{M}_{1}^{e}(P)$ such that 
 $$ \inf _{Y \in \mathscr{A}_{\rho_t}} E_{\tilde{Q}}[Y]>-\infty.$$% Here,  $ \tilde{Q} $ may depend on $t$. 
\end{definition}

\begin{remark}
If $\left\{\rho_{t}\right\}_{0 \leq t \leq T}$ is a dynamic coherent risk measure and continuous from above,   then sensitivity is equivalent to 
$$ P\left(\rho_{t}\left(-\mathbf{1}_{B}\right)>\rho_{t}(0)\right)>0$$
for every $t\in[0,T]$ and  $B\in\mathscr{F}_T$ with $P[B]>0$. This follows from  Lemma 3.4 and Theorem 3.1 in \cite{KS2007} after the sign change $\Phi_{t}=-\rho_{t}$.
\end{remark}
	
The following proposition shows  that a sensitive normalized dynamic star-shaped risk measure can be  represented as the lower envelope of a family of sensitive normalized dynamic convex risk measures.

\begin{proposition}\label{prop:3-3}
Let $\left\{\rho_{t}\right\}_{0 \leq t \leq T}$ be a sensitive normalized dynamic star-shaped risk measure.   Then there exists a nonempty family $\left\{\tilde{\rho}_{t,\lambda} \mid \lambda \in \Lambda\right\}$ of sensitive normalized dynamic convex risk measures  such that
$$\rho_{t}(X)=\mathop{\operatorname{ess}\inf}\limits_{\lambda \in \Lambda} \tilde{\rho}_{t,\lambda}(X)  ,\quad \forall \text X \in L_{T}^{\infty}(P),\quad t\in[0,T],$$and the essential infimum is attained at each fixed time and payoff.
\end{proposition}
\begin{proof}By Proposition \ref{prop:3-2}, it suffices to verify
sensitivity of each constructed component.  In the construction of Proposition \ref{prop:3-2}, for any  $Z \in \mathscr{A}_{t}=\mathscr{A}_{\rho_t}$,  we have   $\mathscr{A}_{t}(Z)\subseteq \mathscr{A}_{t}$ and  $\mathscr{A}_{t}(Z) =\mathscr{A}_{\tilde{\rho}_{t,Z}}$.  Using the sensitivity of $\rho_{t}$,  we obtain for some  $\tilde{Q} \in \mathcal{M}_{1}^{e}(P)$, 
$$ \inf _{X \in \mathscr{A}_{t}(Z)} E_{\tilde{Q}}[X] \geq \inf _{X \in \mathscr{A}_{t}} E_{\tilde{Q}}[X]>-\infty.$$
Hence, each $ \tilde{\rho}_{t,Z} $ is sensitive. \hfill$\Box$
\end{proof}

The same acceptance set inclusion argument, applied to the majorants in Proposition \ref{prop:3-1} and
Corollary \ref{cor:3-1}, yields the following corollary.

\begin{corollary}\label{cor:3-2}
Let $\left\{\rho_{t}\right\}_{0 \leq t \leq T}$  be  a  dynamic monetary (resp. positively homogeneous) risk measure that is sensitive.  Then there exists a nonempty family $\left\{\tilde{\rho}_{t,\lambda} \mid \lambda \in \Lambda\right\}$ of sensitive dynamic convex (resp. coherent) risk measures  such that
$$\rho_{t}(X)=\mathop{\operatorname{ess}\inf}\limits_{\lambda \in \Lambda} \tilde{\rho}_{t,\lambda}(X)  ,\quad \forall \text X \in L_{T}^{\infty}(P), \quad t\in[0,T],$$and the essential infimum is attained at each fixed time and payoff.
\end{corollary}
	
\begin{definition}\label{df:3.2.}
A  dynamic monetary risk measure $\left\{\rho_{t}\right\}_{0 \leq t \leq T}$ is called time consistent  if for every $ X\in L_{T}^{\infty}(P) $,
$$\rho_{t, s}\left(-\rho_{s, T}(X)\right)=\rho_{t, T}(X), \quad \forall 0\leq t\leq s\leq T,$$where $\rho_{t,T}=\rho_t$ and $\rho_{t,s}$ denotes its restriction to $L_s^\infty(P)$.
\end{definition}

The time consistency condition means that one may compute the risk at time $t$ of a financial position with maturity $T$ either directly or in two steps, first at an intermediate time $s$ and then at time $t$ without affecting the result.  For normalized dynamic monetary families, it is equivalent to
$$
\mathscr{A}_{t,T} = \mathscr{A}_{t,s} + \mathscr{A}_{s,T},\quad \forall\,0\le t\le s\le T.
$$
This is an equality of sets and does not require uniqueness of the decomposition. For further discussion of time consistency and acceptance sets, see Lemma 11.14 in \cite{FS16} and Theorem 1 in \cite{BN2009}.  A related question is whether a time consistent dynamic monetary or star-shaped risk measure admits a lower envelope representation by a suitably chosen family of time consistent convex risk measures. Propositions  \ref{prop:3-1} and \ref{prop:3-2}  establish representations at each fixed time, but do not establish time consistency of the representing components.  

We briefly comment on preservation of time consistency under the essential infimum operation. In general, the lower envelope of time consistent dynamic risk measures need not be time consistent. We thank the referee for the following counterexample. Consider four states $\Omega=\{\omega_1,\omega_2,\omega_3,\omega_4\}$ with $P(\omega_i)=1/4$, a trivial $\mathscr F_0$,
$
\mathscr F_1=\sigma\bigl(\{\omega_1,\omega_2\},\{\omega_3,\omega_4\}\bigr)$, $\mathscr F_2=2^\Omega$,
and two equivalent probabilities
$$
Q^1=\frac1{20}(9,1,1,9),
\qquad Q^2=\frac1{20}(1,9,9,1).
$$
Define $\rho_t^i(X)=E_{Q^i}[-X\mid\mathscr F_t]$, $i=1,2$. Both families are normalized, coherent and time consistent. Put $\rho_t(X)=\min\{\rho_t^1(X),\rho_t^2(X)\}$. For $X=(0,-1,0,-1)$,
$$
\rho_0^1(X)=\rho_0^2(X)=\frac12,
\qquad
\rho_1^1(X)=\frac1{10}(1,1,9,9),
\qquad
\rho_1^2(X)=\frac1{10}(9,9,1,1).
$$
Consequently,
$$
\rho_0(X)=\frac12,
\qquad \rho_1(X)=\frac1{10},
\qquad \rho_0\bigl(-\rho_1(X)\bigr)=\frac1{10}\neq\rho_0(X).
$$
Hence,  the lower envelope is not time consistent.  The obstruction is this lack of recursive compatibility, rather than nonlinearity itself.

This example concerns an arbitrary lower envelope, and it does not rule out a suitably chosen time-consistent representing family for a particular risk measure.  In work published after this manuscript was submitted,
\cite{LRM2025} establish related BSDE representation
results in a Lipschitz setting and also study minimal
supersolutions. These results concern frameworks different from the quadratic BSDE setting considered in the next section. Our fixed-time constructions do not establish a corresponding time consistent selection in the quadratic setting. The time consistent BSDE risk measures constructed in Section \ref{sec:4} should therefore be distinguished from the fixed time envelope representations above.

\section{$g$-expectations and dynamic star-shaped risk measures}\label{sec:4}

In this section,  we present several examples of dynamic star-shaped risk measures induced by $g$-expectations. The notion of $g$-expectation was introduced by \cite{P1997} via backward stochastic differential equations (BSDEs; see \cite{PS1990}), and its connection to risk measures has been extensively studied; we refer to  \cite{Rosazza Gianin2006}, \cite{Jiang2005, Jiang2008}, \cite{HMPY2008}, \cite{DPRG2010},  \cite{LRM2025} and the references therein for further details.

Let  $B=\left(B_{t}\right)_{0 \leq t\leq T}$  be a   $d$-dimensional standard Brownian motion defined on a probability space  $(\Omega, \mathscr{F}, P)$  with $B_0 = 0$,  and let  $(\mathscr{F}_{t})_{0 \leq t\leq T}$ be the usual augmented filtration generated by $B$.  % As before, we assume $\mathscr{F}=\mathscr{F}_{T}$,   and for each $t\in[0,T]$,  $L_{t}^{\infty}(P)$ denotes  the space of all essentially bounded $\mathscr{F}_{t}$-measurable random variables.
The scalar generator $g(\omega, t,z): \Omega \times [0,T] \times \mathbb{R}^d \to \mathbb{R}$ is progressively measurable for each $z \in \mathbb{R}^d$.   The following conditions are the basic assumptions on function $g$:
	
(C1) There exists a constant $C>0$ such that, $dP\times dt$-a.e., for any $z\in\mathbb{R}^{d}$, 
$$\left|g(t, z)\right| \leq C\left[1+|z|^{2}\right].$$ 

(C2)  There exists a constant  $K>0$  such that,  $dP\times dt$-a.e., for  any $ z_{1}, z_{2}\in\mathbb{R}^{d}$, 
\begin{align*}
\left|g\left(t, z_{1}\right)-g\left(t, z_{2}\right)\right| \leq K\left(1+|z_{1}|+|z_{2}|\right)|z_{1}-z_{2}|.
\end{align*}
	
(C3) Normalization:  $dP\times dt$-a.e., $g(t, 0)=0$. 
	
(C4) Star-shapedness:  $g$ is star-shaped in $z$: i.e.,  $dP\times dt$-a.e.,
$$\forall \alpha \geq 1, \quad g\left(t, \alpha z\right) \geq \alpha g\left(t, z\right), \quad \text{ for all } z \in \mathbb{R}^{d}. $$
Conditions (C2)--(C3) imply (C1). Unless (C4) is explicitly
imposed, we assume only (C2)--(C3). For each
$\xi\in L_T^\infty(P)$, consider
\begin{equation}\label{equa5}
Y_t=\xi+\int_t^T g(s,Z_s)\,ds
-\int_t^T Z_s\cdot dB_s,\qquad 0\leq t\leq T.
\end{equation}
The quadratic BSDE theory with bounded terminal condition 
provides existence and uniqueness,  see \cite{KM2000}  or \cite{Zhang2017}, Theorem 7.3.3.  We work with continuous bounded adapted solutions $Y$ and predictable $Z$ satisfying
$E_P[\int_0^T|Z_s|^2ds]<\infty$. Following \cite{P1997} and \cite{HMPY2008}, define
$$
\mathcal E_g[\xi\mid\mathscr F_t]:=Y_t(g,T,\xi),
\qquad t\in[0,T],
$$
and write
$\mathcal E_g[\xi]=\mathcal E_g[\xi\mid\mathscr F_0]$.
For related BSDE representations with star-shaped Lipschitz
generators and minimal supersolutions, see \cite{LRM2025}.
Here we consider bounded solutions in the quadratic setting.

\begin{remark}
The comparison theorem applies under (C2) and (C3); see
\cite{Zhang2017}, Theorem 7.3.1. We briefly explain why
comparison is strict at time zero. In fact, for two bounded solutions with the same generator and
terminal values $\xi_1,\xi_2$, set
$\Delta Z=Z^1-Z^2$ and
\[
\beta_s=
\begin{cases}
\displaystyle
\frac{g(s,Z_s^1)-g(s,Z_s^2)}{|\Delta Z_s|^2}\Delta Z_s,
&\Delta Z_s\neq0,\\
0,&\Delta Z_s=0.
\end{cases}
\]
By (C2) and the BMO estimate in \cite{Zhang2017},
Theorem 7.2.1, $\beta\cdot B$ is BMO.
Theorem 7.2.3 therein ensures that its stochastic exponential
defines an equivalent probability $Q$. Under $Q$,
$Y^1-Y^2$ is a bounded martingale, so
\[
Y_t^1-Y_t^2=E_Q[\xi_1-\xi_2\mid\mathscr F_t].
\]
Hence $\xi_1\geq\xi_2$ and $P(\xi_1>\xi_2)>0$ imply
$Y_0^1>Y_0^2$, since $\mathscr F_0$ is trivial and $Q\sim P$.
\end{remark}

\begin{proposition}\label{prop:4-1} 
Suppose  that  the generator $g$ satisfies conditions (C2) and (C3).  Then the following  statements are equivalent.
\begin{itemize}
\item[(i)]  $g$ is star-shaped in $z$.
\item[(ii)] For each $\xi \in L_{T}^{\infty}(P)$,   define
$$\rho_t^{g}(\xi):=\mathcal{E}_{g}\left[-\xi\mid\mathscr{F}_{t}\right],  ~~t\in[0,T].$$
Then $\{\rho_t^{g}(\cdot)\}_{0\leq t\leq T}$ is a normalized time consistent  dynamic star-shaped risk measure. 
\item[(iii)]  For each $\xi \in L_{T}^{\infty}(P)$,  define  $$\rho^{g}(\xi):=\mathcal{E}_{g}\left[-\xi\right].$$Then $\rho^{g}(\cdot)$ is a  normalized  static star-shaped risk measure. 
\end{itemize}
\end{proposition}
	
\begin{proof}   $(ii)\Longrightarrow (iii)$ is  immediate.  $(i)\Longrightarrow (ii)$:  For each constant $\alpha\geq1$, set
$g_\alpha(t,z):=\alpha g(t,z/\alpha).$ 
It also satisfies (C2)-(C3). If $(Y,Z)$ solves the BSDE with generator $g$ and terminal value $\xi$, then $(\alpha Y,\alpha Z)$ solves it with generator $g_\alpha$ and terminal value $\alpha\xi$.
Hence,
$\mathcal E_{g_\alpha}[\alpha\xi\mid\mathscr F_t]
=\alpha\mathcal E_g[\xi\mid\mathscr F_t].$
Since $g$ is star-shaped, $g\geq g_\alpha$. Then it follows from the comparison theorem for BSDE \eqref{equa5} that  for all constants
$\alpha\geq 1$, 
\begin{align}\label{lianxu}\mathcal{E}_{g}\left[\alpha\xi\mid\mathscr{F}_{t}\right]\geq\alpha\mathcal{E}_{g}\left[\xi\mid\mathscr{F}_{t}\right], ~~\forall t\in[0,T],~~ \xi\in L^{\infty}_{T}(P).\end{align}
By uniqueness of BSDE and condition (C3),  one obtains that for any  $A\in \mathscr{F}_{t}$,  
\begin{align}\label{eq:14}
\mathcal{E}_{g}\left[I_{A}\xi\mid\mathscr{F}_{t}\right]=I_{A}\mathcal{E}_{g}\left[\xi\mid\mathscr{F}_{t}\right], ~ \quad \xi\in L^{\infty}_{T}(P).
\end{align}
Consequently,  \eqref{lianxu} holds for all simple functions $\alpha\in L^{\infty}_{t}(P)$ and $\alpha\geq 1$.  By the continuous dependence on the terminal data (Theorem~2.8 in \cite{KM2000} or Theorem~7.3.4 in \cite{Zhang2017}), the inequality \eqref{lianxu}  remains valid for all  $\alpha\in L_t^\infty(P)$ with $\alpha\ge 1$. 
Thus  $\rho_t^{g}(\cdot)$ is star-shaped. The time consistency, normalization and the monetary property are obvious.  Hence $\{\rho_t^{g}(\cdot)\}_{0\le t\le T}$ is a normalized time consistent dynamic star-shaped risk measure.

$(iii)\Longrightarrow (i)$.  By the star-shapedness of  $\rho^{g}(\cdot)$, we have that for all constants $\alpha \geq 1$, 
\begin{align}\label{eq:15}
\mathcal{E}_{g}\left[\alpha\xi\right]\geq \alpha \mathcal{E}_{g}\left[\xi\right],~~\forall \xi\in L^{\infty}_{T}(P).
\end{align} We now show that \eqref{lianxu} holds for every constant $\alpha\ge 1$. To this end, set
 $$A:=\{\mathcal{E}_{g}\left[\alpha\xi\mid\mathscr{F}_{t}\right]<\alpha\mathcal{E}_{g}\left[\xi\mid\mathscr{F}_{t}\right]\}.$$Then $A\in\mathscr{F}_{t}$. Suppose, for contradiction, that $P(A)>0$.  By the basic properties of BSDE~\eqref{equa5} and  the strict comparison established above we obtain
 \begin{align*}
\mathcal{E}_{g}\left[I_{A}\alpha\xi-I_{A}\alpha\mathcal{E}_{g}\left[\xi\mid\mathscr{F}_{t}\right]\right]&=\mathcal{E}_{g}\left[\mathcal{E}_{g}\left[I_{A}\alpha\xi-I_{A}\alpha\mathcal{E}_{g}\left[\xi\mid\mathscr{F}_{t}\right]\mid\mathscr{F}_{t}\right]\right]\\
&=\mathcal{E}_{g}\left[I_{A}\Big(\mathcal{E}_{g}\left[\alpha\xi\mid\mathscr{F}_{t}\right]-\alpha\mathcal{E}_{g}\left[\xi\mid\mathscr{F}_{t}\right] \Big)\right]<0.
\end{align*}
On the other hand,  using the star-shapedness of $\mathcal{E}_{g}[\cdot]$ together with \eqref{eq:14},   we deduce that
\begin{align*}
\mathcal{E}_{g}\left[I_{A}\alpha\xi-I_{A}\alpha \mathcal{E}_{g}\left[\xi\mid\mathscr{F}_{t}\right]\right]&\geq  \alpha \mathcal{E}_{g}\left[I_{A}\xi-I_{A}\mathcal{E}_{g}\left[\xi\mid\mathscr{F}_{t}\right]\right]\\
&=\alpha\mathcal{E}_{g}\left[\mathcal{E}_{g}\left[I_{A}\xi-I_{A}\mathcal{E}_{g}\left[\xi\mid\mathscr{F}_{t}\right]\mid\mathscr{F}_{t}\right]\right]\\
&=\alpha\mathcal{E}_{g}\left[I_{A}\Big(\mathcal{E}_{g}\left[\xi\mid\mathscr{F}_{t}\right]-\mathcal{E}_{g}\left[\xi\mid\mathscr{F}_{t}\right] \Big)\right]=0.\end{align*}This is a contradiction.   Therefore, $P(A)=0$ and  inequality \eqref{lianxu} holds for all constants $\alpha\geq 1$. 
Furthermore, by the scaling identity, one has
$$
\mathcal E_g[\xi\mid\mathscr F_t]
\geq\alpha\mathcal E_g[\xi/\alpha\mid\mathscr F_t]
=\mathcal E_{g_\alpha}[\xi\mid\mathscr F_t].
$$

To apply a converse comparison or representation theorem,  fix a stopping time $\tau\leq T$
and $\xi\in L_\tau^\infty(P)$. For either generator
$h\in\{g,g_\alpha\}$, let $(Y,Z)$ be the bounded solution on
$[0,\tau]$, whose existence follows by stopping the generator.
Define
$$
\widetilde Y_t=Y_{t\wedge\tau},
\qquad
\widetilde Z_t=1_{\{t\leq\tau\}}Z_t.
$$
Since $h(t,0)=0$, this extension solves the BSDE on $[0,T]$
with generator $h$ and terminal value $\xi$. By uniqueness, the preceding comparison holds on $[0,\tau]$.  Applying it at rational times and using continuity makes it valid simultaneously at all times.

Both generators satisfy the assumptions of \cite{Zheng2015},
Corollary 3.3, since they are independent of the value variable,
continuous in $z$, and satisfy a deterministic quadratic
growth bound. Thus, for each fixed $\alpha\geq1$ and
$z\in\mathbb R^d$,
$$
g(t,z)\geq g_\alpha(t,z)=\alpha g(t,z/\alpha),
\qquad dP\times dt\text{-a.e.}
$$
Fubini's theorem, a countable dense set argument and continuity
give a common null set outside which this inequality holds
for all $\alpha\geq1$ and $z\in\mathbb R^d$.
Replacing $z$ by $\alpha z$ proves (i).
\hfill$\Box$
\end{proof}

\begin{corollary}\label{cor:4-1}
Suppose the generator $g$ satisfies conditions (C2)-(C4).  Then $\{\rho_t^{g}(\cdot)\}_{0\leq t\leq T}$ is a normalized time consistent  dynamic star-shaped risk measure on $L^{\infty}_{T}(P)$, and for each $t$, there exists a nonempty family $\left\{\rho_{t,\lambda} \mid \lambda \in \Lambda\right\}$ of continuous from above, normalized dynamic convex risk measures such that
$$\rho_t^{g}(\xi)=\mathop{\operatorname{ess}\inf} _{\lambda \in \Lambda} \rho_{t,\lambda}(\xi),\quad \forall \xi  \in L_{T}^{\infty}(P).$$	The essential infimum is attained at each fixed time and payoff.
\end{corollary}

We conclude with several examples of dynamic risk measures from BSDEs. To infer failures of convexity and positive homogeneity from the generators, we first introduce the following lemma.

\begin{lemma}\label{lem:auxi}
Suppose the generator $g$ satisfies conditions (C2)-(C3). If $\rho_t^g$ is conditionally convex
(resp. conditionally concave or conditionally positively
homogeneous) for every $t\in[0,T]$, then $g$ has the
corresponding property in $z$ outside a common
$dP\times dt$-null set. \end{lemma}

\begin{proof}
Since $\rho_t^g(\xi)=\mathcal E_g[-\xi\mid\mathscr F_t]$,
it suffices to consider $\mathcal E_g$. For $t<T$, set
$$
\sigma_\varepsilon=(t+\varepsilon)\wedge T\wedge
\inf\{s\geq t:|B_s-B_t|\geq1\},
\qquad
\eta_z^\varepsilon=z\cdot(B_{\sigma_\varepsilon}-B_t).
$$
Then $|\eta_z^\varepsilon|\leq|z|$. By the quadratic growth bound and the stopped horizon extension in the proof of Proposition \ref{prop:4-1},
\cite{Zheng2015}, Corollary 3.2(ii), yields, for each fixed
$z\in\mathbb R^d$ and almost every $t<T$,
$$
g(t,z)=\lim_{\varepsilon\downarrow0}
\frac{\mathcal E_g[\eta_z^\varepsilon\mid\mathscr F_t]}
{\varepsilon},
\qquad P\text{-a.s.}
$$
Since $\sigma_\varepsilon$ is independent of $z$,
$$
\eta_{bz_1+(1-b)z_2}^\varepsilon
=b\eta_{z_1}^\varepsilon+(1-b)\eta_{z_2}^\varepsilon,
\qquad
\eta_{az}^\varepsilon=a\eta_z^\varepsilon.
$$Applying the corresponding property of $\mathcal E_g$,
dividing by $\varepsilon$ and passing to the limit yields
the desired property of $g$. Fubini's theorem, a countable
dense-set argument and continuity in $z$ give a common
product-measure null set.
\hfill $\Box$
\end{proof}

\begin{example}For each $z\in\mathbb{R}$, define 
$$g(z)=|z|^{4} I_{|z| \leq 1}+|z|^{2} I_{|z|>1}. $$
This generator satisfies (C2)-(C3). For $r>0$, $g(r)/r$ equals $r^3$ on $(0,1]$ and $r$ on $(1,\infty)$; it is increasing, so $g$ is star-shaped. It is not convex, since its left and right derivatives at $1$ are $4$ and $2$. It is not concave, since $g(0)<(g(-h)+g(h))/2$ for $0<h<1$, and not positively homogeneous, since $g(2)\ne2g(1)$. By Proposition \ref{prop:4-1} and Lemma \ref{lem:auxi}, it indicates that the induced family is normalized, time consistent and star-shaped, but neither conditionally convex nor conditionally concave nor conditionally positively homogeneous.
\end{example}
	
\begin{example}($\alpha$-maxmin expectations) 
Maxmin expectations are widely used in economics, finance and decision theory as a tractable class of nonlinear expectations.  Fix $\kappa>0$ and let
$$
\Gamma=\left\{Q^\theta:\frac{dQ^\theta}{dP}
 =\exp\left(\int_0^T\theta_s\cdot dB_s
 -\frac12\int_0^T|\theta_s|^2ds\right),\quad |\theta_s|\leq\kappa\right\},
$$
where $\theta$ ranges over predictable $\mathbb R^d$-valued processes. The bounded drifts ensure that each $Q^\theta$ is a probability
measure equivalent to $P$. Following \cite{CK2006}, define
$$
\overline{\mathcal E}[\xi\mid\mathscr F_t]
 =\operatorname{ess\,sup}_{Q\in\Gamma}E_Q[\xi\mid\mathscr F_t],
\qquad
\underline{\mathcal E}[\xi\mid\mathscr F_t]
 =\operatorname{ess\,inf}_{Q\in\Gamma}E_Q[\xi\mid\mathscr F_t].
$$
These are the BSDE evaluations with generators $\kappa|z|$ and $-\kappa|z|$, respectively. For  $\alpha\in[0,1]$, let
$$
\mathcal E^\alpha[\xi\mid\mathscr F_t]
 =\alpha\overline{\mathcal E}[\xi\mid\mathscr F_t]
 +(1-\alpha)\underline{\mathcal E}[\xi\mid\mathscr F_t].
$$
Both extreme evaluations preserve constants, monotonicity,  translation invariance and multiplication by nonnegative bounded $\mathscr F_t$-measurable factors. Hence, 
$\rho_t^\alpha(\xi):=\mathcal E^\alpha[-\xi\mid\mathscr F_t]$
is a normalized dynamic positively homogeneous risk measure, and therefore star-shaped.  Positive homogeneity alone guarantees neither convexity nor time consistency of the weighted family.  For dynamically consistent $\alpha$-maxmin constructions, see \cite{BLR20}.
\end{example}		
	
\begin{example}(Entropic risk measures with drift ambiguity)
Consider the following BSDE: 
\begin{align}\label{shangfengxian}
			Y_{t}=-\xi+\int_{t}^{T} \big(g(Z_{s})+\frac{\gamma}{2}|Z_{s}|^{2}\big)ds-\int_{t}^{T} Z_{s} \cdot d B_{s}, \quad 0 \leq t \leq T,
\end{align}where $\xi\in L^{\infty}_{T}(P)$ and $\gamma>0$. 
%The generator of BSDE \eqref{shangfengxian} is $$f(z)=g(z)+\frac{\gamma}{2}|z|^{2}, \quad z\in\mathbb{R}.$$
If $g$ satisfies (C2)-(C3) and is positively homogeneous, then,  using the exponential transform $U=e^{\gamma Y}$, $V=\gamma UZ$ and It\^{o}'s formula, one gets
$$
dU_s=-\gamma U_sg(Z_s)ds+V_s\cdot dB_s
 =-g(V_s)ds+V_s\cdot dB_s,\qquad U_T=e^{-\gamma\xi}.
$$
Since $Y$ is bounded, then $U$ is bounded and $V\cdot B$ is BMO. Hence, it yields that 
\begin{align}
\label{eq:quadratic}
\rho_{t}(\xi):=Y_t=\frac{1}{\gamma} \ln \mathcal{E}_{g}[e^{-\gamma\xi } | \mathscr{F}_{t}],\quad \xi\in L^{\infty}_{T}(P).%&=\frac{1}{\gamma} \ln \mathop{\operatorname{ess}\inf} _{Q \in \Gamma} E_{Q}\left[e^{-\gamma\xi} \mid \mathscr{F}_{t}\right]
%&=\mathop{\operatorname{ess}\inf} _{Q \in \Gamma}  \frac{1}{\gamma} \ln E_{Q}\left[e^{\gamma\xi} \mid \mathscr{F}_{t}\right].
\end{align}
For $g=0$, it reduces to the classical dynamic entropic risk measure.

For instance, in dimension one, we take
$$
g(z)=\kappa_1z^+-\kappa_2z^-
 =\inf_{\theta\in[\kappa_1,\kappa_2]}\theta z,
 \qquad 0<\kappa_1<\kappa_2.
$$
Let $\Theta$ contain all predictable processes taking values in $[\kappa_1,\kappa_2]$, and define $\Gamma=\{Q^\theta:\theta\in\Theta\}$ by
$$
\frac{dQ^\theta}{dP}
 =\exp\left(\int_0^T\theta_s\,dB_s
 -\frac12\int_0^T|\theta_s|^2ds\right).
$$The bounded drifts ensure that these densities define equivalent
probabilities. By Girsanov's theorem, it then implies
$$
dU_s=(\theta_sV_s-g(V_s))ds+V_s\cdot dB_s^{Q^\theta},
 \qquad \theta_sV_s-g(V_s)\geq0.
$$
Thus,  $U$ is a bounded submartingale under every $Q^\theta$, by localization, and
$U_t\leq E_{Q^\theta}[e^{-\gamma\xi}\mid\mathscr F_t]$.
Equality holds for taking
$\theta_s^*=\kappa_1 1_{\{V_s\geq0\}}+\kappa_2 1_{\{V_s<0\}}$, 
under which $U$ is a bounded martingale.  Therefore, 
$$
\rho_t(\xi)=\operatorname{ess\,inf}_{\theta\in\Theta}
 \frac1\gamma\log E_{Q^\theta}[e^{-\gamma\xi}\mid\mathscr F_t],
$$
with attainment at $\theta^*$. This represents an optimistic choice of the smallest entropic risk evaluation across drift models. Time consistency follows from the BSDE flow property, with predictable drifts allowing
model choices to adapt over time.

Finally, the generator $f(z)=g(z)+\gamma|z|^2/2$ of
BSDE \eqref{shangfengxian} satisfies (C2)-(C3) and is
star-shaped, since
$$
f(\alpha z)-\alpha f(z)
=\frac{\gamma}{2}\alpha(\alpha-1)|z|^2\geq0,
\qquad \alpha\geq1.
$$
It is not positively homogeneous, and is neither convex nor
concave: for $0<\varepsilon<(\kappa_2-\kappa_1)/\gamma$,
\[
\frac{f(\varepsilon)+f(-\varepsilon)}2<0=f(0),
\]
whereas $f$ is strictly convex on $(0,\infty)$.
By Proposition \ref{prop:4-1} and Lemma \ref{lem:auxi}, the induced family
$\{\rho_t\}_{t\in[0,T]}$ is normalized, time consistent and star-shaped,
but neither conditionally convex nor conditionally concave
nor conditionally positively homogeneous.
\end{example}

\section{Conclusions}\label{sec:5}

Motivated by \cite{JXZ2020} and \cite{CCMW2022}, 
this paper establishes representation theorems for dynamic risk measures. Analogous to their static counterparts, dynamic monetary risk measures admit attained lower envelope representations by dynamic convex risk measures at each fixed time, and normalized dynamic star-shaped risk measures admit such representations with normalized convex components. The positively homogeneous case yields coherent majorants without requiring continuity from above or countably additive dual representations. Furthermore, we characterize fixed time translations between monetary and star-shaped risk measures through an essentially bounded common acceptance level, and examine sensitivity and time consistency. Several examples, together with a characterization of normalized time consistent star-shaped risk measures induced by quadratic $g$-expectations under (C2)-(C3), are presented. The fixed time representations do not establish time consistency of the representing components; selecting time consistent convex components in the quadratic BSDE setting remains a direction for further research.

\section*{Declaration of Competing Interests}

The authors declare that they have no known competing financial interests or personal relationships that could have appeared to influence the work reported in this paper. 

\section*{Acknowledgments}
The authors acknowledge financial support from the National
Natural Science Foundation of China (No. 12671583).
The first author thanks Dr. Shiqiu Zheng for helpful discussions
on the converse comparison theorem.
The authors also thank the editor, associate editor, and referee
for their valuable comments and suggestions, which have
significantly improved the manuscript.

\section*{Declaration of generative AI and AI-assisted technologies in the manuscript preparation process}
During the preparation of this work, the authors used Deepseek and ChatGPT for assisting with language editing and checking mathematical arguments. The authors reviewed and edited the output as needed and take full responsibility for the content of the published article.

%\section*{References}


\begin{thebibliography}{99}
\linespread{0.9}\selectfont % 行内间距

	\bibitem[Artzner et al.(1999)]{ADEH1999}Artzner, P., Delbaen, F., Eber, J.M., Heath, D. 1999. Coherent measures of risk. Mathematical Finance, 9(3), 203-228.	
				
		\bibitem[Artzner et al.(2007)]{APDFEJHK2007}Artzner, P., Delbaen, F., Eber, J.M., Heath, D., Ku, H. 2007. Coherent multi-period risk adjusted values and Bellman’s principle. Annals of Operations Research, 152, 5-22.

     \bibitem[Beissner et al.(2020)]{BLR20} Beissner, P., Lin, Q., Riedel, F. 2020. Dynamically consistent alpha-maxmin expected utility.  Mathematical Finance,  30(3), 1073-1102. 		
		
    \bibitem[Bion-Nadal (2008)]{BN2008}Bion-Nadal, J. 2008. Dynamic risk measures: time consistency and risk measures from BMO martingales. Finance and Stochastics, 12(2), 219-244.
		
		\bibitem[Bion-Nadal (2009)]{BN2009}Bion-Nadal, J. 2009. Time consistent dynamic risk processes. Stochastic Processes and their Applications, 119(2), 633-654.
		
\bibitem [Castellani (2000)]{C2000} Castellani, M. 2000. A dual representation for proper positively homogeneous functions. Journal of Global Optimization, 16, 393-400.
			    \bibitem[Castagnoli et al.(2022)]{CCMW2022}Castagnoli, E., Cattelan, G., Maccheroni, F., Tebaldi, C., Wang, R. 2022. Star-shaped risk measures. Operations Research, 70, 2637-2654.

	    	
\bibitem [Castagnoli et al.(2015)]{CDFM2015}Castagnoli, E., De  Donno, M., Favero, G.,  Modesti, P. 2015. Granular and star-shaped price systems. Journal of Financial Risk Management, 4, 227-249.

    

		\bibitem [Chen and Kulperger (2006)]{CK2006}Chen, Z., Kulperger, R. 2006. Minimax pricing and Choquet pricing. Insurance:  Mathematics and Economics, 38(3), 518-528.
				
		\bibitem [Delbaen and Schachermayer (1994)]{DS1994}Delbaen, F., Schachermayer, W. 1994. A general version of the fundamental theorem of asset pricing. Mathematische Annalen, 300(1), 463-520.
		
		\bibitem[Delbaen et al.(2010) ]{DPRG2010}Delbaen, F., Peng, S., Rosazza Gianin, E. 2010. Representation of the penalty term of dynamic concave utilities. Finance and Stochastics, 14(3), 449-472.
		
		\bibitem[Detlefsen and Scandolo (2005) ]{DS2005}Detlefsen, K., Scandolo, G. 2005. Conditional and dynamic convex risk measures. Finance and Stochastics, 9(4), 539-561.
       
		\bibitem[F{\"o}llmer and Penner (2006)]{FP2006}F{\"o}llmer, H., Penner, I. 2006. Convex risk measures and the dynamics of their penalty functions. Statistics \& Risk Modeling, 24(1), 61-96.
			
		\bibitem[F\"{o}llmer and Schied (2002)]{FS2002}F\"{o}llmer, H., Schied, A. 2002. Convex measures of risk and trading constraints. Finance and Stochastics, 6(4), 429-447.
				
		\bibitem[F\"{o}llmer and Schied (2016)]{FS16}F\"{o}llmer, H., Schied, A. 2016. Stochastic Finance: An Introduction in Discrete Time, 4th Edition. De Gruyter Studies in Mathematics, Berlin, Germany.
		
\bibitem[Frittelli and Rosazza Gianin (2002)]{FRG2002}Frittelli, M., Rosazza Gianin, E. 2002. Putting order in risk measures. Journal of Banking \& Finance, 26(7), 1473-1486.
	\bibitem[Frittelli and Rosazza Gianin (2004)]{FRG2004}Frittelli, M., Rosazza Gianin, E. 2004. Dynamic convex risk measures. Risk Measures for the 21st Century, 227-248.
		
		\bibitem[Hu et al.(2008)]{HMPY2008}Hu, Y., Ma, J., Peng, S., Yao, S. 2008. Representation theorems for quadratic $ \mathscr F$-consistent nonlinear expectations. Stochastic Processes and their Applications, 118(9), 1518-1551.
					
\bibitem[Ji et al.(2019)]{Ji2019}Ji, R.,  Shi, X.,  Wang, S.,  Zhou, J. 2019.  Dynamic risk measures for processes via backward stochastic differential equations. Insurance: Mathematics and Economics, 86, 43-50.					
\bibitem[Jia et al.(2020)]{JXZ2020}Jia, G., Xia, J., Zhao, R. 2020. Monetary risk measures. arXiv preprint arXiv:2012.06751.
				
\bibitem[Jiang (2005)]{Jiang2005}Jiang, L. 2005. Representation theorems for generators of backward stochastic differential equations and their applications. Stochastic Processes and their Applications, 115(12), 1883-1903.
		
\bibitem[Jiang (2008)]{Jiang2008}Jiang, L. 2008. Convexity, translation invariance and subadditivity for $ g $-expectations and related risk measures. The Annals of Applied Probability, 18(1), 245-258.
						
\bibitem[Kobylanski (2000)]{KM2000}Kobylanski, M. 2000. Backward stochastic differential equations and partial differential equations with quadratic growth. The Annals of Probability, 28(2), 558-602.
		
\bibitem[Kl\"{o}ppel and Schweizer (2007)]{KS2007}Kl\"{o}ppel, S., Schweizer, M. 2007. Dynamic indifference valuation via convex risk measures. Mathematical Finance, 17(4), 599-627.

\bibitem[Laeven and Rosazza Gianin (2022)]{LR2022}Laeven, R.J.A., Rosazza Gianin, E. 2022. Quasi-logconvex measures of risk. arXiv:2208.07694v1.

\bibitem[Laeven  et al.(2024)]{LRZ2024} Laeven, R.J.A., Rosazza Gianin, E., Zullino, M. 2024. Law-invariant return and star-shaped risk measures. Insurance: Mathematics and Economics, 117, 140-153.


\bibitem[Laeven et al.(2026)]{LRM2025}Laeven, R.J.A., Rosazza Gianin, E.,  Zullino, M. 2026.  Star-shaped and dynamic return risk measures via BSDEs. Finance and Stochastics, 30, 903-950.


%\bibitem[Mao and Wang (2020)]{MW2020}Mao, T., Wang, R. 2020. Risk aversion in regulatory capital principles. SIAM Journal on Financial Mathematics, 11(1), 169-200.
		
\bibitem[Moresco and Righi (2022)]{MR2022}Moresco, M. R., Righi, M. B.  2022. On the link between monetary and star-shaped risk measures. Statistics \& Probability Letters, 184, 109345.
		
\bibitem[Natarajan et al.(2008)]{NPS2008}Natarajan, K., Pachamanova, D., Sim, M. 2008. Incorporating asymmetric distributional information in robust value-at-risk optimization. Management Science, 54(3), 573-585.

\bibitem [Nie et al.(2025)]{NTJ2025}
        Nie, B., Tian, D., Jiang, L. 2025. Set-valued star-shaped risk measures. Mathematics and Financial Economics, 19, 329-350.
 

		
\bibitem[Pardoux and Peng (1990)]{PS1990}Pardoux, E., Peng, S. 1990. Adapted solution of a backward stochastic differential equation. Systems \& Control Letters, 14(1), 55-61.
		
\bibitem[Peng (1997)]{P1997}Peng, S. 1997. Backward SDE and related $ g $-expectation. Pitman Research Notes in Mathematics Series, 141-160.


				
\bibitem[Riedel (2004)]{Riedel2004}Riedel, F. 2004. Dynamic coherent risk measures. Stochastic Processes and their Applications, 112(2), 185-200.

\bibitem[Righi (2024)]{Righi2022}Righi, M.B. 2024. Star-shaped acceptability indexes. Insurance: Mathematics and Economics, 117, 170-181.

		
\bibitem[Rosazza Gianin (2006)]{Rosazza Gianin2006}Rosazza Gianin, E. 2006. Risk measures via g-expectations. Insurance: Mathematics and Economics, 39(1), 19-34.		
			
\bibitem[Wang and Ziegel (2021)]{WZ2021}Wang, R., Ziegel, J. F. 2021. Scenario-based risk evaluation. Finance and Stochastics, 25, 725-756.
						
\bibitem[Zhang (2017)]{Zhang2017}Zhang, J. 2017. Backward Stochastic
Differential Equations: From Linear to Fully Nonlinear Theory.  Springer, New York, NY. 
	
\bibitem[Zheng and Li (2015)]{Zheng2015}Zheng, S., Li, S. 2015. Representation theorems for generators of BSDEs with monotonic and convex growth generators. Statistics \& Probability Letters, 97, 197-205.
		
		
\end{thebibliography}
\end{document}